\documentclass[preprint]{aastex631}

\shorttitle{Sulfur monoxide as a tracer of the Galactic $^{32}$S/$^{34}$S gradient}
\usepackage{CJK}
\usepackage{hyperref}
\usepackage{longtable}
\usepackage{multirow}
\usepackage{threeparttable}
\usepackage{pgffor}
\usepackage{subfigure}
\usepackage{amsmath}
\usepackage{float}
\setcitestyle{aysep={}}
\usepackage{enumitem}
\setlist{parsep=0pt, itemsep=10pt, partopsep=0pt} 

\begin{document}
\begin{CJK*}{UTF8}{gbsn}
\title{Galactic Interstellar $^{32}$S/$^{34}$S Isotopic Ratio traced by Sulfur Monoxide}
\author[0000-0002-5230-8010]{Y. P. Zou (邹益鹏)}
\affiliation{Center for Astrophysics, Guangzhou University, 
Guangzhou, 510006, People's Republic of China}
\affiliation{Department of Physics and Astronomy, University of Bologna, Via Gobetti 93/2, 40129 Bologna, Italy}
\affiliation{INAF, Astrophysics and Space Science Observatory Bologna, Via Gobetti 93/3, 40129 Bologna, Italy}

\author[0000-0002-5161-8180]{J. S. Zhang (张江水)}
\affiliation{Center for Astrophysics, Guangzhou University, 
Guangzhou, 510006, People's Republic of China}
\correspondingauthor{J. S. Zhang}
\email{jszhang@gzhu.edu.cn}

\author{D. Y. Wei (韦丁元)}
\affiliation{Center for Astrophysics, Guangzhou University, 
Guangzhou, 510006, People's Republic of China}

\author[0000-0001-5574-0549]{Y. T. Yan (闫耀庭)}
\affiliation{Max-Planck-Institut f{\"u}r Radioastronomie, Auf dem H{\"u}gel 69, D-53121 Bonn, Germany}

\author[0000-0002-0845-6171]{D. Romano}
\affiliation{INAF, Astrophysics and Space Science Observatory Bologna, Via Gobetti 93/3, 40129 Bologna, Italy}

\author[0000-0001-9155-0777]{Y. X. Wang (汪友鑫)}
\affiliation{Center for Astrophysics, Guangzhou University, 
Guangzhou, 510006, People's Republic of China}
\affiliation{Max-Planck-Institut f{\"u}r Radioastronomie, Auf dem H{\"u}gel 69, D-53121 Bonn, Germany}

\author[0000-0001-8980-9663]{J. L. Chen (陈家梁)}
\affiliation{School of Mathematics and Physics, Jinggangshan University, Jiangxi, 343009, People's Republic of China}

\author[0000-0002-5634-131X]{H. Z. Yu (余鸿智)}
\affiliation{Ural Federal University, 19 Mira Street, 620002 Ekaterinburg, Russia}

\author{J. Y. Zhao (赵洁瑜)}
\affiliation{Center for Astrophysics, Guangzhou University, 
Guangzhou, 510006, People's Republic of China}

\begin{abstract}         
To date, the Galactic interstellar radial $^{32}$S/$^{34}$S gradient has only been studied with the CS isotopologs, which may be affected by uncertainties due to the use of a single tracer. As another abundant S-bearing molecules, SO and its isotopomer $^{34}$SO could be considered as tracers of the $^{32}$S/$^{34}$S ratio. We present the first systematic observations of SO and $^{34}$SO toward a large sample of molecular clouds with accurate distances, performed with the IRAM 30 m and the 10 m Submillimeter Telescope (SMT). With the IRAM 30 m, SO $2_2-1_1$ was detected in 59 of 82 sources ($\sim$82\%), and $^{34}$SO $2_2-1_1$ in 8 sources ($\sim$10\%). With the SMT 10 m, SO $5_5-4_4$ was detected in 136 of 184 sources ($\sim$74\%), and $^{34}$SO $5_5-4_4$ in 55 of 77 strong SO sources ($\sim$72\%). SO/$^{34}$SO ratios were derived for 8 ($2_2-1_1$) and 55 ($5_5-4_4$) sources. No correlation was found between the SO/$^{34}$SO ratio and heliocentric distance or $T_k$, suggesting negligible distance and fractionation effects. Both LTE and non-LTE analyses consistently suggest that the optical depth effect is also insignificant. $^{32}$S/$^{34}$S ratios from the $2_2-1_1$ transitions follow the increasing radial trend proposed by previous CS species measurements, while those from the $5_5-4_4$ lines are systematically lower. The lower transitions of SO and $^{34}$SO may be suitable tracers of $^{32}$S/$^{34}$S, though the detections are rare. Comparisons between measurements and Galactic chemical evolution model suggest that the nucleosynthesis prescriptions need to be revised in the low-metallicity regime, but more data for the outermost Galactic regions are crucial for drawing strong conclusions.
\end{abstract}

\section{Introduction} \label{sec:intro}
Stellar nucleosynthesis converts hydrogen into heavier elements, subsequent stellar ejections change the chemical composition of the interstellar medium (ISM), driving the so-called chemical evolution of galaxies. Isotope abundance ratios provide a powerful tool for tracing stellar nucleosynthesis and constraining the Galactic chemical evolution (GCE; e.g., \citealt{Wilson1994}; \citealt{2008A&A...487..237W}; \citealt{2015ApJS..219...28Z}, \citeyear{Zhang2020}; \citealt{Romano2022}). These ratios can be effectively determined through observations of molecular clouds, specifically by analyzing the molecular lines of corresponding isotopologues. Isotope ratios, such as $^{12}$C/$^{13}$C (e.g., \citealt{Yan2019, Yan2023}), $^{14}$N/$^{15}$N (e.g., \citealt{Chen2021, Colzi2022, Chen2024}), $^{18}$O/$^{17}$O (e.g., \citealt{Zhang2020, Zou2023}) and $^{32}$S/$^{34}$S (e.g., \citealt{Chin1996, Yu2020, Yan2023}), are systematically measured in star-forming regions (SFRs) across the Galactic disc.

Of these isotopes, carbon, nitrogen, and oxygen are mainly produced via hydrogen, helium, and carbon burning processes (e.g., \citealt{Wilson1994, Romano2022}), while sulfur and its stable isotopologues provide unique insights into the chemical enrichment by oxygen-, silicon-, and neon-burning, as well as s-process nucleosynthesis (e.g., \citealt{Gong2023}). Both $^{32}$S and its isotopologues ($^{34}$S, $^{33}$S, and $^{36}$S) predominantly originate from massive stars, with a minor contribution from asymptotic giant branch (AGB) stars in the case of $^{36}$S \citep{Pignatari2016}. Specifically, the synthesis of $^{32}$S is linked to oxygen burning through nuclear fusion of $^{28}$Si and $^{4}$He \citep{Hughes2008}, $^{34}$S and $^{33}$S are produced as byproducts of pre-explosive oxygen burning through neutron capture on $^{32}$S \citep{Pignatari2016}, while the abundance of $^{36}$S is predominantly associated with the weak s-process \citep{Mauersberger2004}. These isotopes are ejected into the ISM during the explosive death of massive stars, such as core-collapse (Type II, Ib, and Ic) supernovae (e.g., \citealt{Timmes1995, Kobayashi2020}).Only a minor fraction of $^{36}$S is ejected through AGB winds \citep[][and references therein]{Mauersberger2004}.

Sulfur isotope ratios in the ISM across the Galactic disc are systematically measured by observations of CS and its isotopologues, and a positive radial gradient of $^{32}$S/$^{34}$S ratio was reported \citep{Chin1996, Yu2020, Yan2023}. However, the determination of $^{32}$S/$^{34}$S ratios in these studies mostly relies on the double isotope method ($^{13}$CS/C$^{34}$S $\times$ $^{12}$C/$^{13}$C) due to the high opacity of CS. Although \cite{Yan2023} also derived the $^{32}$S/$^{34}$S ratios directly from $^{13}$CS/$^{13}$C$^{34}$S, only 17 sources were successfully measured due to the weakness of the line. The double isotope method may introduce additional uncertainties due to the uncertain Galactic $^{12}$C/$^{13}$C gradient. On the other hand, the direct measurements using $^{13}$CS/$^{13}$C$^{34}$S may be limited to strong sources and thus cannot provide reliable statistical results. Furthermore, relying on a single tracer to measure the ratio may introduce systematic errors. Through comparisons of ratios measured from different molecular species, systematic discrepancies related to the chosen molecular species can be assessed. Thus it is necessary to explore other suitable tracers for systematic measurement of sulfur isotope ratios.
 
After CS, SO is the most abundant and easily observable S-bearing molecule in the gas phase \citep{Rodriguez-Baras2021}. The SO/$^{34}$SO ratio was proposed as one of the indicators for the $^{32}$S/$^{34}$S ratio in several previous studies. However, these studies focused on individual sources, such as the well-known Sgr B2 and Orion KL \citep{Nummelin2000, Esplugues2013}. Here, we present the first systematic observations of SO and $^{34}$SO toward a large sample of molecular clouds with accurate distances, to explore whether the SO species can effectively serve as suitable tracers for systematic measurements of the $^{32}$S/$^{34}$S ratio. The observations with the IRAM 30 m and the SMT 10 m are described in Section \ref{sec:obs}. The data reduction and our results are presented in Section \ref{sec:d&r}, while Section \ref{sec:a&d} contains an analysis and the discussion. Our main findings are summarized in Section \ref{sec:sum}.

\section{Observations}\label{sec:obs}
\subsection{Sample information}
Over the past decade, mainly thanks to the Bar and Spiral Structure Legacy (BeSSeL) Survey\footnote{http://bessel.vlbi-astrometry.org}, the trigonometric parallaxes of approximately 200 high-mass star-forming regions (HMSFRs) have been determined across the Milky Way through dedicated high resolution observations of molecular maser lines \citep{Reid2014, Reid2019}. This provides a good sample with reliable distances, since the trigonometric parallax method can determine the distance to the sources very directly and accurately. 191 among them were observed in this work, due to the limitation in sky coverage during our observations. In addition, 15 sources in the outer disk ($R_{\mathrm{GC}}$ $>$ 8 kpc, selected from \citeauthor{Wouterloot1989} \citeyear{Wouterloot1989}) are added to our sample, to cover the entire Galactic disk (see Table \ref{tab:lineparameter} for the total sample). The SO and $^{34}$SO $2_2-1_1$ line data used in this work were obtained from our previous IRAM 30 m projects, which focused on the $^{18}$O/$^{17}$O ratio (see \citealt{Zhang2020} and \citealt{Zou2023}). For sources with ratio measurements, we estimated the kinematic distances from the Revised Kinematic Distance calculator \citep{Reid2014} for two sources (G034.30+00.15, WB89 171) without trigonometric parallax data.
\subsection{The IRAM 30 m observations}
With the IRAM 30 m telescope, we observed the $2_2-1_1$ transitions of SO and $^{34}$SO toward 82 sources. Our observations were carried out in 2016 June and July (project 013-16) and 2017 January (project 088-16). Using FTS200 backend with a resolution of 195 kHz ($\sim$ 0.68 km s$^{-1}$), the observations were performed in total power position switching mode and the off position was set at 30$\arcmin$ in R.A. and Dec. (or azimuth). The antenna temperature $T_A^*$ can be transformed into the main beam brightness temperature $T\rm_{mb}$, through the relation $T\rm_{mb}$ = $T_A^*$ $\times$($F_{eff}/B_{eff}$)\footnote{https://publicwiki.iram.es/Iram30mEfficiencies}. The half power beam width (HPBW) for each line was calculated by HPBW($\arcsec$)=2460/$\nu$(GHz). Rest frequencies, excitations of the upper levels above the ground state, Einstein coefficients for spontaneous emission, spectral resolutions on velocity scale after smoothing, telescope efficiencies, the RMS ranges and corresponding median values, as well as beam sizes are listed in Table \ref{tab:obsparameters}. Sixty of these 82 sources were also observed by the SMT 10 m.
\subsection{The SMT 10 m observations}
Through the SMT 10 m telescope, the observations for SO and $^{34}$SO $5_5-4_4$ lines were performed remotely in 2021 April and May (project Zou\_21A\_1), 2022 April and May (projects Zou\_22A\_1, 2). We first carried out a survey of the SO $5_5-4_4$ line toward 184 sources, then further observed the $^{34}$SO $5_5-4_4$ line for the strongest 77 sources ($T_{A}$(SO) $\gtrsim$ 0.3 K). The dual-polarization 1.3 mm Receiver frontends and the SMT Filter banks backends with 2 IF mode were used, which provide bandwidths of 1000 and 256 MHz and spectral resolutions of 1000 and 250 kHz ($\sim$ 1.40 and 0.35 km s$^{-1}$), respectively. Position switching mode with the off position of 30$\arcmin$ in azimuth was adopted. On the antenna temperature scale $T_A^*$, the system temperature was $\sim$ 221 K with an rms noise of $\sim$ 12 mK in our observations. The main beam brightness temperature $T\rm_{mb}$ can be derived from the antenna temperature, via the equation $T\rm_{mb}$ = $T_A^*$/$\eta_{mb}$\footnote{https://aro.as.arizona.edu/?q=beam-efficiencies}. The HPBW was estimated by $\lambda/D$ with a 10 metre diameter ($D$). Detailed observational information are presented in Table \ref{tab:obsparameters}. 

\begin{table*}[htpb]
 \centering
 \caption{Information on targeted spectral lines of SO and $^{34}$SO.}
 \setlength{\tabcolsep}{1.5mm}{
  \begin{tabular}{ccccccccccc}
  \hline\hline
  Telescope & Isotopologue & Transition & $\nu_0$ & $E_{u}/k$ & $A_{u, l}$ & $g_u$ & $\Delta V$ & $C_{eff}$ & RMS & HPBW\\
  & & $N_J-N'_{J'}$ & (GHz) & (K) & (s$^{-1}$) &  & (km s$^{-1}$)& & (mK) & ($\arcsec$)\\
   (1) & (2) & (3) & (4) & (5) & (6) & (7) & (8) & (9) & (10) & (11)\\
   \hline
   \multirow{2}{*}{ IRAM 30 m} & SO & $2_2-1_1$ & 86.093958 & 19.3 & $5.25\times10^{-6}$ & 5 & 1.36 & \multirow{2}{*}{1.17} & 9-22 (14) & 28.6\\
 & $^{34}$SO & $2_2-1_1$ & 84.410684 & 19.2 & $4.95\times10^{-6}$ & 5 & 1.38 &  & 8-24 (14) & 29.1\\
   \hline
   \multirow{2}{*}{ SMT 10 m} & SO & $5_5-4_4$ & 215.220650 & 44.1 & $1.19\times10^{-4}$ & 11 & 1.39 & \multirow{2}{*}{1.41} & 11-57 (17) & 28.8\\
 & $^{34}$SO & $5_5-4_4$ & 211.013019 & 43.5 & $1.12\times10^{-4}$ & 11 & 1.42 & & 3-31 (9) & 29.3\\
   \hline
  \end{tabular}}
 \tablecomments{Column (1): telescope, column (2): species, column (3): the quantum number of the line transition, column (4): rest frequency, column (5): upper energy level, column (6): Einstein coefficient for spontaneous emission from upper $u$ to lower $l$ level, column (7): the statistical weight of the upper state, column (8): spectral resolution on velocity scale, column (9): correction factor for telescope efficiency, column (10): the RMS range and corresponding median value on $T_{\rm mb}$ scale at the final velocity resolution, column (11): half power beam width. ($A_{ul}$, $g_u$, $E_u$ were collected from the CDMS database, \citealt{Muller2005, Endres2016} and $\nu_{0}$ was taken from \citealt{Esplugues2013})}
 \label{tab:obsparameters}
\end{table*}

\section{Data reduction and results}\label{sec:d&r}
In order to improve the signal-to-noise ratio (SNR), the SMT 10 m data with a lower spectral resolution (1000 kHz; $\sim$ 1.40 km s$^{-1}$) were finally used. The IRAM 30 m data were smoothed for comparison with the SMT 10 m data (see Table \ref{tab:obsparameters} for the final spectral resolution). The data of both telescopes were reduced using the Continuum and Line Analysis Single-dish Software (CLASS) of the Grenoble Image and Line Data Analysis Software packages (GILDAS\footnote{https://www.iram.fr/IRAMFR/GILDAS/}). The baseline was subtracted using polynomial fitting, and the parameters of the detected lines (SNR $>$ 3) were obtained through Gaussian fitting. We noticed that some sources exhibit SO line wing, which may trace regions different from that of the central narrow component. \cite{Esplugues2013} performed good fits of SO lines of the Orion KL with five known components. Then they further generated synthetic spectra of SO and $^{34}$SO using the LVG model, and obtained the corresponding SO/$^{34}$SO ratios for these components. However, unlike the well studied Orion KL, it is not clear how many distinct components are in our sources, and thus making it difficult to reliably fit the $^{34}$SO line with multiple components. If we only fit SO lines with multiple components, it is still hard to make appropriate line width comparisons, and to derive the SO/$^{34}$SO ratios corresponding to the narrow and the wide component. Therefore, we favor a single Gaussian fit unless both SO and $^{34}$SO clearly show consistent multiple components. Spectral line parameters of SO and $^{34}$SO measured by both the IRAM 30 m and the SMT 10 m are presented in Table \ref{tab:lineparameter}.
\setlength{\tabcolsep}{1.5mm}{
\begin{deluxetable*}{ccccccccc}
\tabletypesize{\footnotesize}
\tablecaption{\label{tab:lineparameter} Spectral line parameters of SO and $^{34}$SO lines observed by the IRAM 30 m and the SMT 10 m.}
\tablehead{
\colhead{Source} & \colhead{R.A.} & \colhead{Decl.} & \colhead{Line} & \colhead{RMS} & \colhead{$v_{LSR}$} & \colhead{$\Delta v_{1/2}$} & \colhead{$\int T\rm_{mb}d{\nu }$} &
\colhead{$T\rm_{mb}$}\\
\colhead{} & \colhead{(J2000)} & \colhead{(J2000)} & \colhead{} & \colhead{(mK)} & \colhead{(km s$^{-1}$)} & \colhead{(km s$^{-1}$)} & \colhead{(K km $s^{-1}$)} & \colhead{(K)}\\
\colhead{(1)} & \colhead{(2)} & \colhead{(3)} & \colhead{(4)} & \colhead{(5)} &
\colhead{(6)} & \colhead{(7)} & \colhead{(8)} &
\colhead{(9)}
}
\startdata
        WB89 330 & 00 20 58.10 & 62 40 18.01 & SO ($2_2-1_1$) & 11 & ... & ... & ... & ... \\
        ~ & ~ & ~ & $^{34}$SO ($2_2-1_1$) & 14 & ... & ... & ... & ... \\
        WB89 331 & 00 21 19.40 & 63 19 19.99 & SO ($2_2-1_1$) & 14 & ... & ... & ... & ... \\
        ~ & ~ & ~ & $^{34}$SO ($2_2-1_1$) & 14 & ... & ... & ... & ... \\
        G121.29+00.65 & 00 36 47.35 & 63 29 02.15 & SO ($5_5-4_4$) & 18 & -18.30 $\pm$ 0.05 & 3.32 $\pm$ 0.14 & 1.90 $\pm$ 0.06 & 0.54 \\
        ~ & ~ & ~ & $^{34}$SO ($5_5-4_4$) & 7 & -17.3 $\pm$ 0.4 & 4.8 $\pm$ 1.0 & 0.15 $\pm$ 0.03 & 0.03 \\
        WB89 354 & 00 38 59.46 & 59 27 48.82 & SO ($2_2-1_1$) & 12 & ... & ... & ... & ... \\
        ~ & ~ & ~ & $^{34}$SO ($2_2-1_1$) & 13 & ... & ... & ... & ... \\
        G122.01-07.08 & 00 44 58.40 & 55 46 47.60 & SO ($5_5-4_4$) & 20 & -51.62 $\pm$ 0.17 & 1 $\pm$ 8 & 0.15 $\pm$ 0.04 & 0.10 \\
        \multicolumn{9}{c}{...}\\
\enddata
\tablecomments{Column (1): source name, columns (2) and (3): equatorial coordinates (J2000), column (4): molecular species and its transition, column (5): the rms noise value, column (6): LSR velocity, column (7): line width (FWHM), column (8): integrated line intensity; column (9): peak $T\rm_{mb}$ value.\\
(This table is available in its entirety in machine-readable form.)}
\end{deluxetable*}
}

\begin{figure*}[htpb]
\centering
\includegraphics[width=\textwidth, page=1]{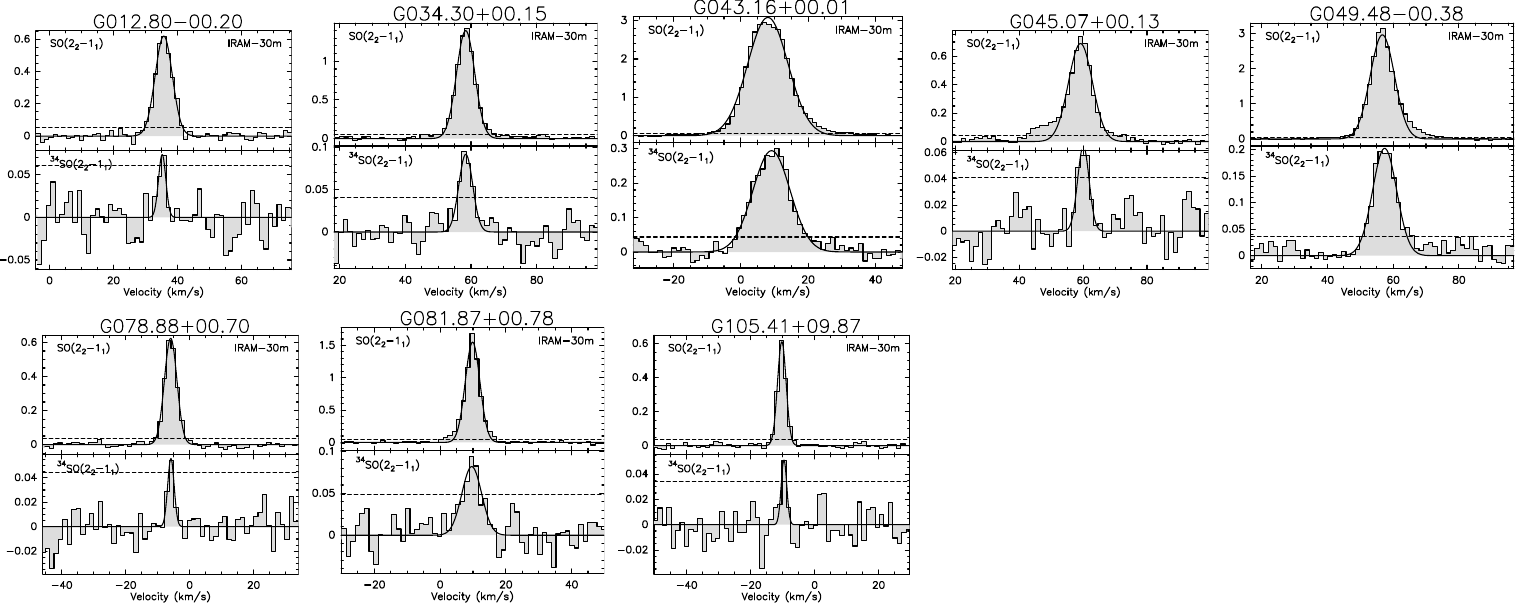}
\caption{The IRAM 30 m spectra with Gaussian fit line of SO $2_2-1_1$ (upper panels) and $^{34}$SO $2_2-1_1$ (lower panels) of our detected sources. The dashed horizontal line indicates the 3$\sigma$ level.}
\label{fig:both2-1}
\end{figure*}

Our analysis reveals that, out of 82 sources observed with the IRAM 30 m, 59 were detected in SO $2_2-1_1$ and 8 were detected in $^{34}$SO $2_2-1_1$. Among the 184 sources observed with the SMT 10 m, 136 were successfully detected in SO $5_5-4_4$, while 55 were detected in $^{34}$SO $5_5-4_4$, starting from a subsample of 77 strong SO sources ($T_{A}$(SO) $\gtrsim$ 0.3 K). Thus we obtained the intensity ratio of SO/$^{34}$SO for 8 and 55 sources from $2_2-1_1$ and $5_5-4_4$ transitions data, respectively, including 7 sources measured in both transitions. The spectra of these sources, together with Gaussian fits, are presented in Figure \ref{fig:both2-1} (IRAM 30 m) and Figure \ref{fig:both5-4} (SMT 10 m), respectively. Their physical parameters, including the intensity ratio, the heliocentric distance, and the galactocentric distance are listed in Table \ref{tab:ratio}. The spectra of the sources without effective ratio measurements are shown in Appendix \ref{sec:appendF}.
 
Large variations are apparent among line widths of different sources, e.g., the line width of G000.67-00.03 is greater than 25 km s$^{-1}$, while that of G111.23-01.23 is less than 4 km s$^{-1}$. To investigate the cause of these variations, we plotted the line width of SO and $^{34}$SO against the galactocentric distance for our sample in Figure \ref{fig:vSO_34SO_rgc}. We found a negative correlation between the line width and the galactocentric distance, with correlation coefficients of -0.59 and -0.48 for SO and $^{34}$SO, respectively. In other words, the line width decreases as the galactocentric distance increases. This could be related to more active star formation in the inner Galaxy than in the outer region, which would lead to more turbulence and gas motions (but the line broadening mechanisms can be complex, see a short discussion at the end of Section. \ref{sec:cp}).

\begin{figure*}[htpb]
\centering
\includegraphics[width=\textwidth, page=1]{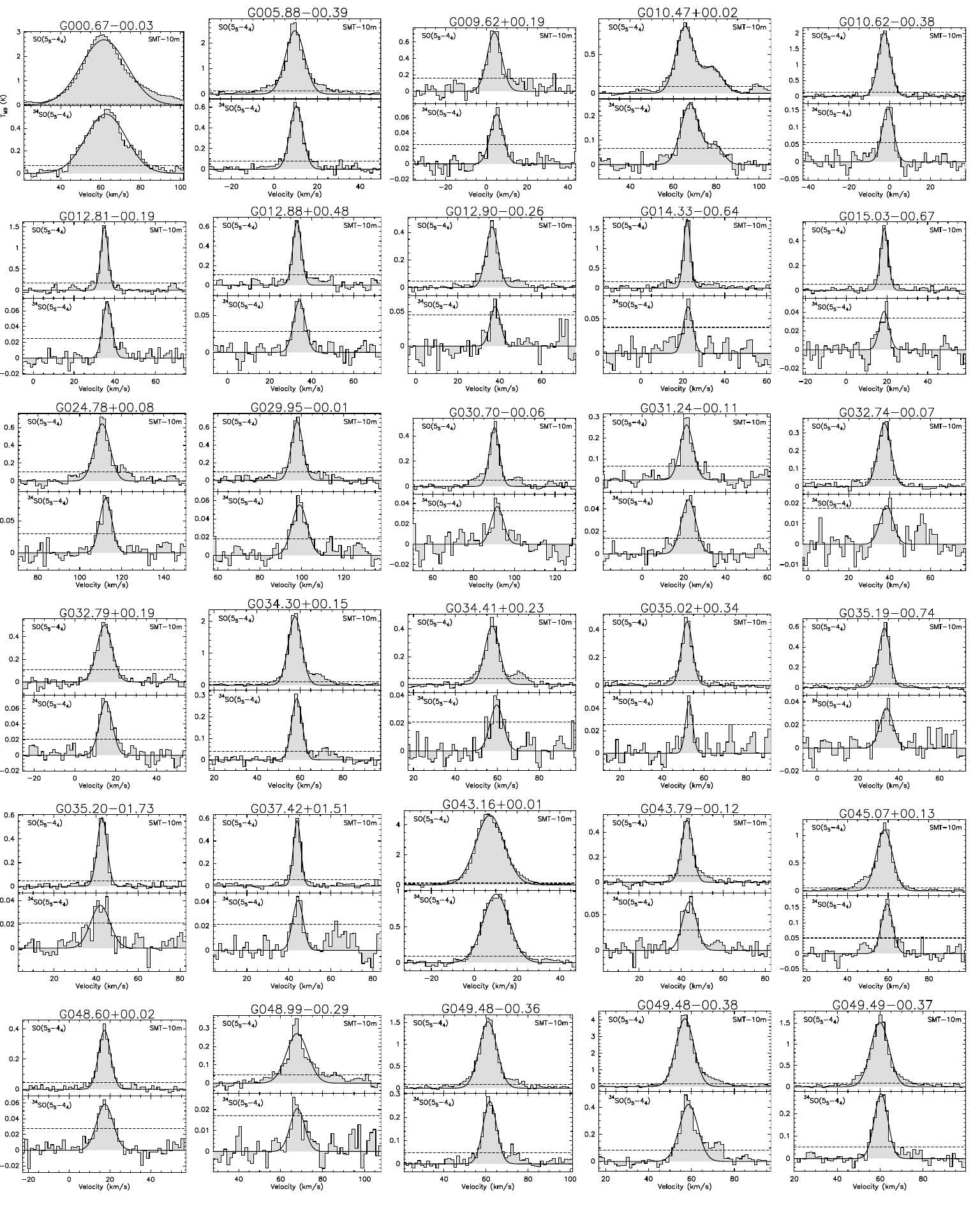}
\caption{The SMT 10 m spectra of SO $5_5-4_4$ (upper panels) and $^{34}$SO $5_5-4_4$ (lower panels) with fit lines of the 55 sources in both isotopomers. The dashed horizontal lines indicate the 3$\sigma$ levels.}
\label{fig:both5-4}
\end{figure*}
\addtocounter{figure}{-1}
\begin{figure*}[ht]
\includegraphics[width=\textwidth, page=1]{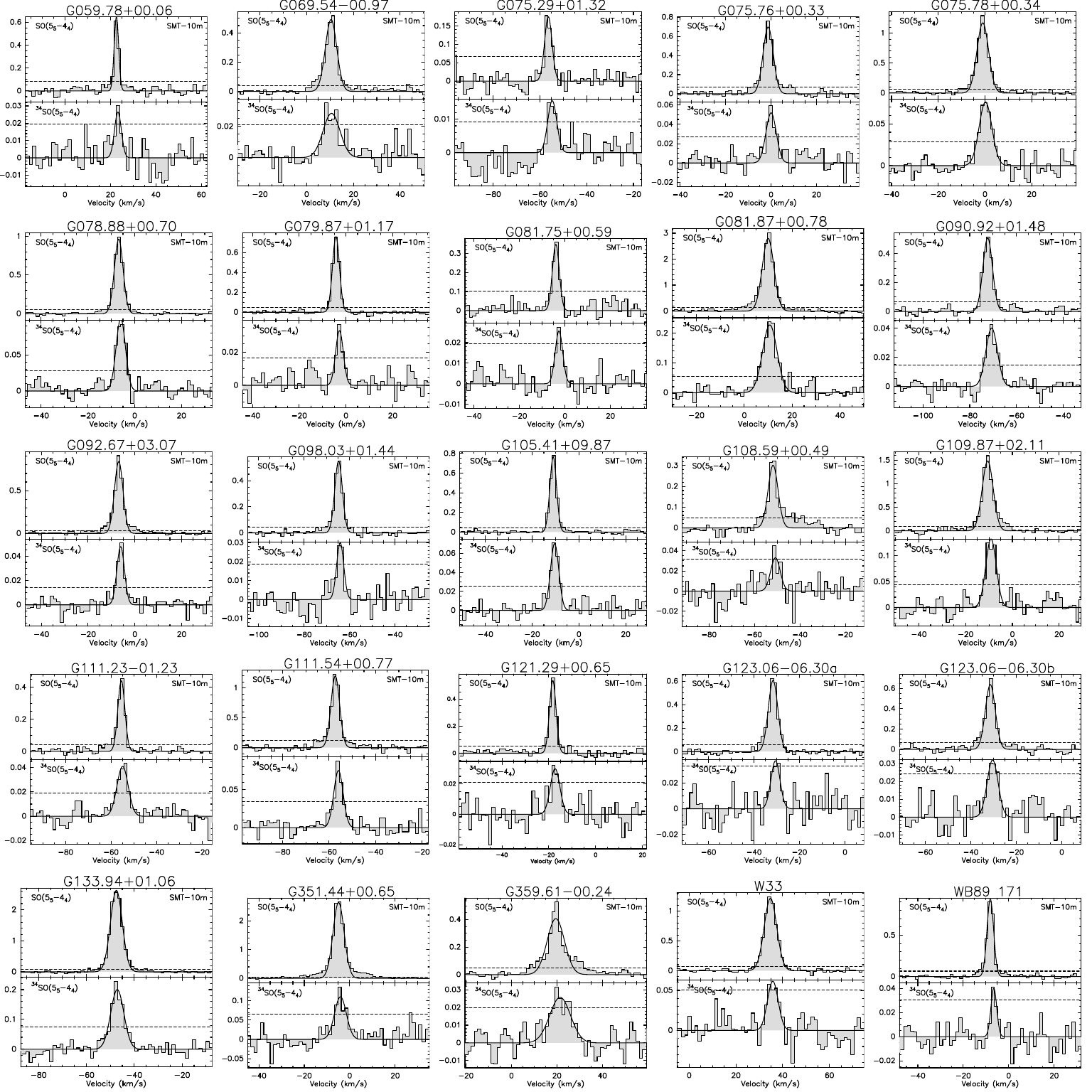}
\caption{continued.}
\end{figure*}

\begin{figure}[htpb]
\centering
\includegraphics[width=\textwidth]{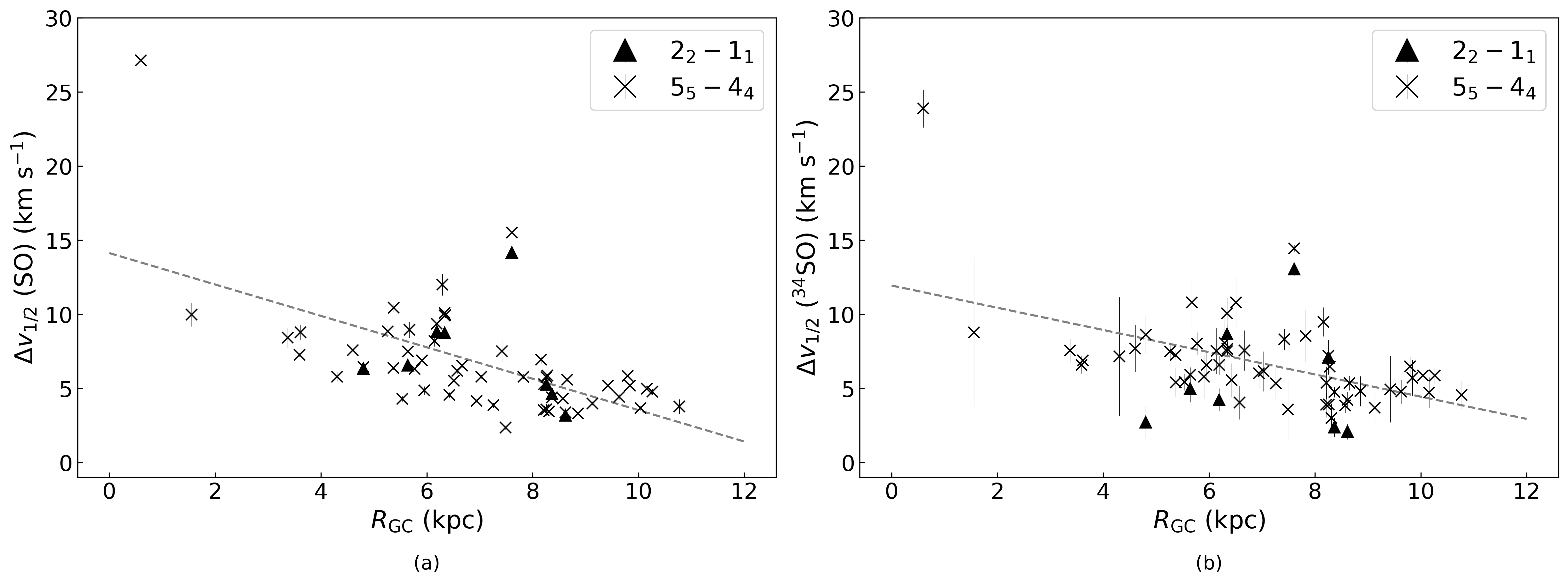}
\caption{The line width of SO (a) and $^{34}$SO (b) as a function of the galactocentric distance for sources detected in both lines. The linear fit result is plotted as the gray dashed line.}
\label{fig:vSO_34SO_rgc}
\end{figure}

\section{Analysis and Discussion}\label{sec:a&d}
\subsection{Possible effects on abundance ratios}
In the following, we discuss possible physical mechanisms that might skew the observed isotopogue abundance ratios from the isotopic ones.
\subsubsection{Observational effects}
Two antennas with different diameters, the IRAM 30 m and the SMT 10 m, were used for our observations. This could result in a beam area difference of approximately one order of magnitude at the same targeted frequency. Taking this into consideration, the IRAM 30 m telescope was primarily used to observe lower transition lines, while the SMT 10 m telescope was employed for higher transitions. As a result of this strategy, the beam size is basically consistent for the lines targeted with the two telescopes (28.6-29.3$\arcsec$; see Table \ref{tab:obsparameters}). Thus, there is no significant effect on our ratio results.

\begin{figure}[htbp]
\centering
\includegraphics[width=0.6\textwidth]{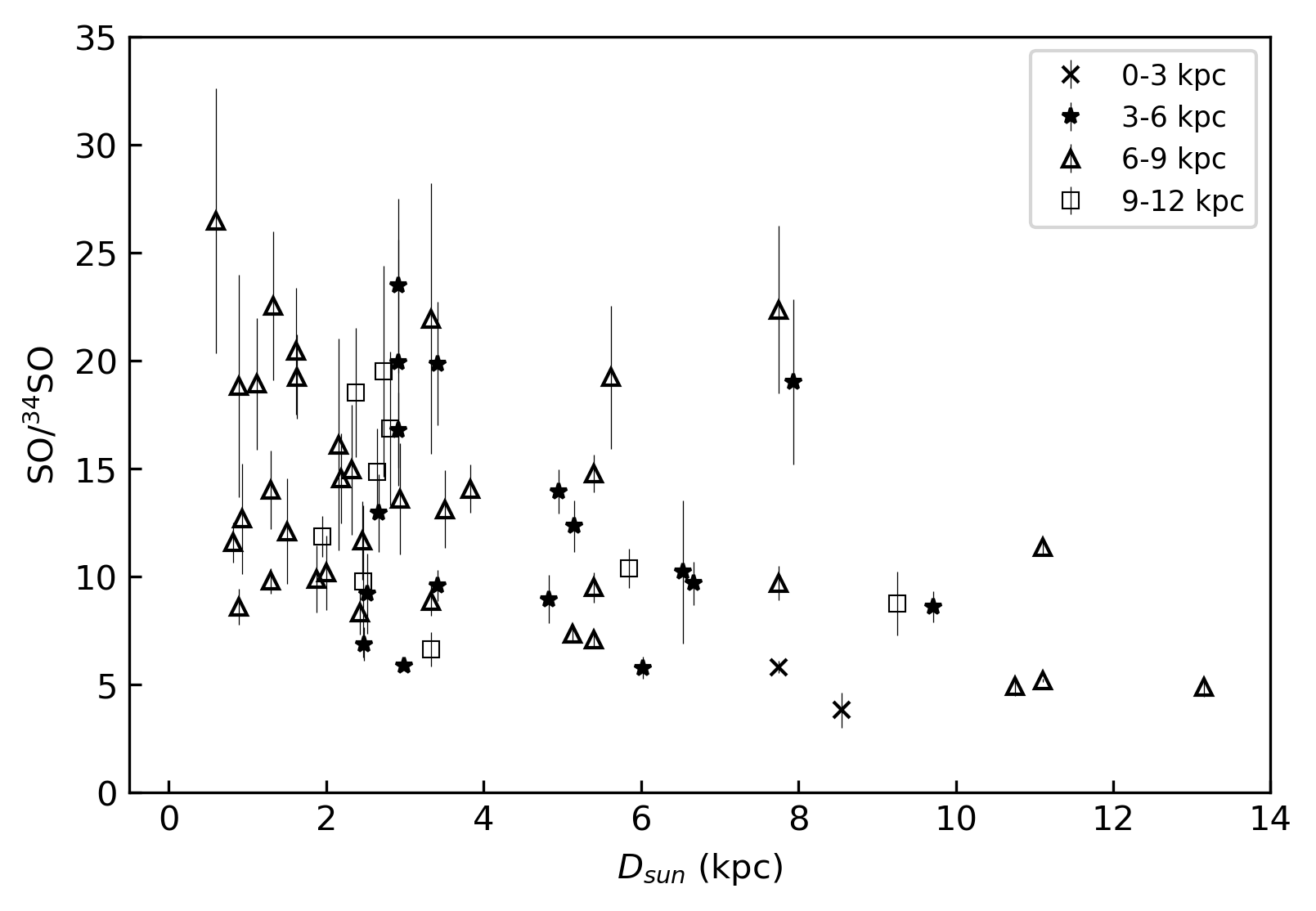} 
\caption{Intensity ratio of SO/$^{34}$SO against the heliocentric distance. As the ratio may vary with $R_{\mathrm{GC}}$, it is better to divide sources in different $R_{\mathrm{GC}}$ bins when analyzing other factors. The crosses, stars, triangles, and squares represent sources in different radial galactocentric bins.}
\label{fig:r_d}
\end{figure}

There might still be present some biases related to the distance from the Sun ($D_{sun}$), such as the effect of different linear resolutions and a bias to brighter sources at larger distances. To evaluate these potential effects on our ratio results, we plotted the isotopic ratio against the heliocentric distance in Figure \ref{fig:r_d}. No systematic dependence was found between the ratio and heliocentric distance. Hence, any distance dependent effects are considered insignificant.

\subsubsection{Chemical fractionation}
\begin{figure}[htbp]
\centering
\includegraphics[width=0.6\textwidth]{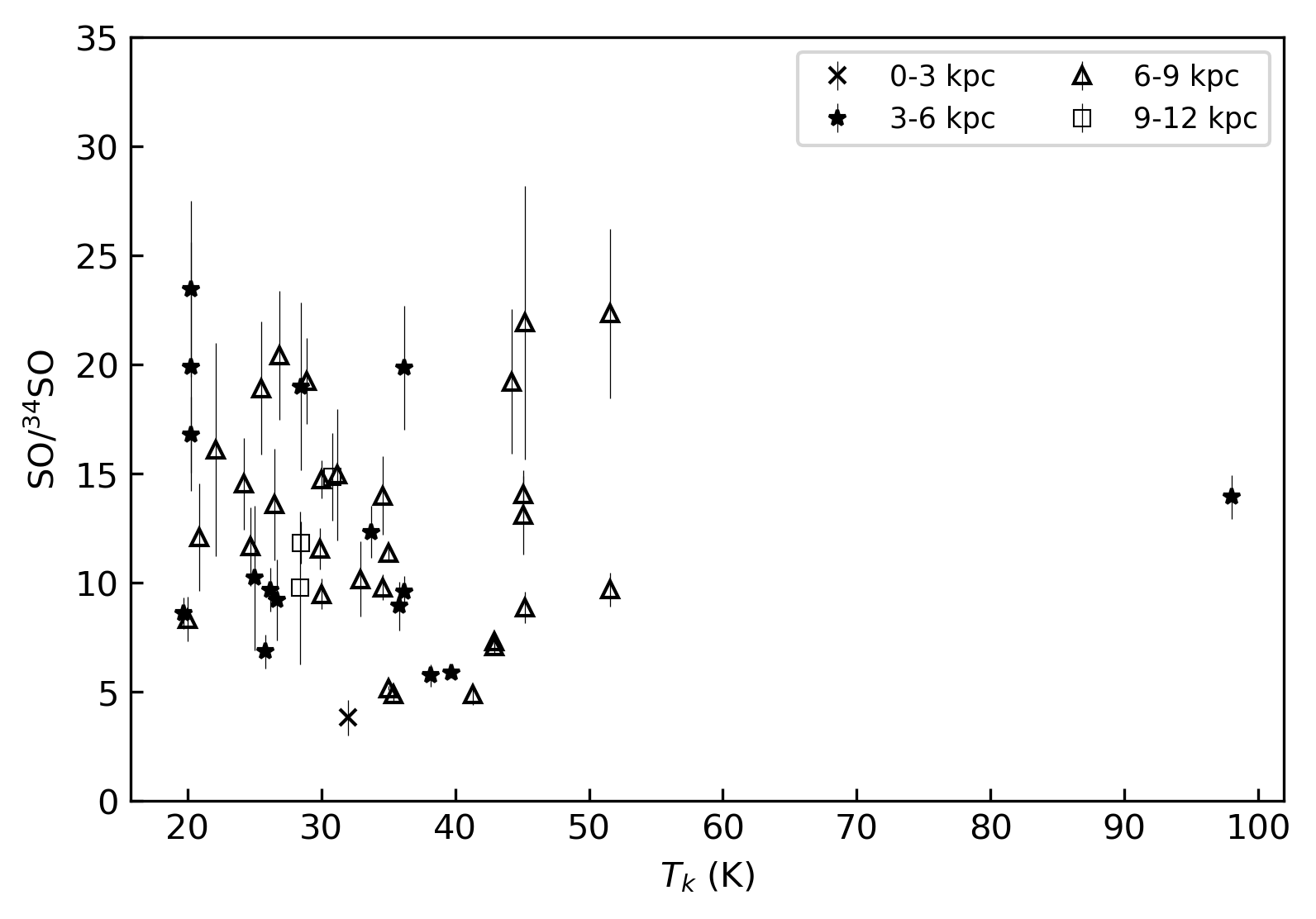} 
\caption{Intensity ratio of SO/$^{34}$SO against the gas kinetic temperature. The crosses, stars, triangles, and squares represent sources in different radial galactocentric bins.}
\label{fig:r_Tk}
\end{figure}

The observed isotope ratios may not accurately reflect stellar nucleosynthesis and chemical evolution if they are significantly influenced by fractionation \citep{Chin1996}. Previous theoretical calculations and related observational data suggest that sulfur fractionation processes are negligible in the cold cores of dense clouds \citep{Loison2019}, as well as in warm clouds of the Galactic center \citep{Humire2020}. To further investigate possible fractionation effects toward our sample, we plotted the SO/$^{34}$SO ratio against the gas kinetic temperature, $T_{k}$, for different radial galactocentric bins in Figure \ref{fig:r_Tk}. The kinetic temperatures for our sources were collected from several previous studies (see Table \ref{tab:ratio} for references), in which they were estimated using the para-NH3 (1, 1) and (2, 2) transitions. No correlation was found between the SO/$^{34}$SO ratio and $T_{k}$, indicating that fractionation effects related to the temperature do not play an important role. 

\subsubsection{Optical depth}\label{sec:tau}
As mentioned in the introduction, until now the Galactic radial gradient of $^{32}$S/$^{34}$S has only been estimated with the CS isotopologs. Relying on a single tracer to measure the ratio may introduce systematic errors. Actually, the $^{32}$S/$^{34}$S ratio can also be directly determined by using sulfur monoxide and its isotopologue $^{34}$SO. It should be noted, however, that the use of optically thin tracers or the correction for optical depth is important for properly estimating the column densities and isotopic ratios. Unlike the rare isotopologues ($^{13}$CS, C$^{34}$S, and $^{13}$C$^{34}$S) used in the previous CS measurements (e.g., \citealt{Yan2023}), SO is more abundant and thus may not be optically thin in some cases. In this section, we employ two methods (LTE and non-LTE) to estimate the optical depth of SO and evaluate its potential effect on our ratio results. Two additional SO lines ($3_2-2_1$ and $5_6-4_5$) were observed with the IRAM 30 m (projects 004-20 and 125-20; Zou et al. in preparation) and combined to our previous observations to estimate the optical depths.

\setlength{\tabcolsep}{1mm}{
\setlength{\extrarowheight}{2mm}
\startlongtable
\begin{deluxetable*}{cccccccccccc}
\tabletypesize{\scriptsize}
\tablecaption{ Physical parameters of sources with measured isotope ratio SO/$^{34}$SO.\label{tab:ratio}}
\tablehead{
\colhead{Transition} & \colhead{Source} & \colhead{$D_{sun}$} & \colhead{$R_{\rm GC}$} & \colhead{I(SO)/I($^{34}$SO)} & \colhead{$\tau_{\rm SO}$(LTE)} & \colhead{$\tau_{\rm SO}$(nLTE)} & \colhead{$T_{k}$} & \colhead{$T_{rot}$} & \colhead{$N_{rot}$} & \colhead{$T_{\rm LTE}$} & \colhead{$N_{\rm LTE}$}\\
\colhead{} & \colhead{} & \colhead{(kpc)} & \colhead{(kpc)} & \colhead{} & \colhead{} & \colhead{} & \colhead{(K)} & \colhead{(K)} & \colhead{($\times$10$^{14}$ cm$^{-2}$)} & \colhead{(K)} & \colhead{($\times$10$^{14}$ cm$^{-2}$)}\\
 \colhead{(1)} & \colhead{(2)} & \colhead{(3)} & \colhead{(4)} & \colhead{(5)} &\colhead{(6)} & \colhead{(7)} & \colhead{(8)} & \colhead{(9)} & \colhead{(10)} & \colhead{(11)} & \colhead{(12)}\\
 }
\startdata
$2_2-1_1$ & G012.80-00.20  & 2.92  & 4.80  & 20 $\pm$ 6     & ...  & ...  & 20.3$^{(h)}$ & 47 $\pm$ 40 & 4 $\pm$ 3     & ...             & ...                    \\
$2_2-1_1$ & G034.30+00.15  & 10.58 & 5.70  & 20 $\pm$ 3     & 0.12 & 0.10 & 36.2$^{(g)}$ & 18 $\pm$ 8  & 5.6 $\pm$ 1.4 & $17^{+5}_{-3}$  & $7.2^{+1.5}_{-1.4}$    \\
$2_2-1_1$ & G043.16+00.01  & 11.11 & 7.60  & 11.4 $\pm$ 0.3 & 0.23 & 0.21 & 35.0$^{(b)}$ & 17 $\pm$ 5  & 26 $\pm$ 6    & $18^{+6}_{-4}$  & $35^{+9}_{-7}$         \\
$2_2-1_1$ & G045.07+00.13  & 7.75  & 6.19  & 22 $\pm$ 4     & 0.05 & 0.05 & 51.6$^{(h)}$ & 19 $\pm$ 26 & 4 $\pm$ 2     & $16^{+6}_{-3}$  & $4.3^{+1.1}_{-1.0}$    \\
$2_2-1_1$ & G049.48-00.38  & 5.41  & 6.34  & 14.8 $\pm$ 0.9 & 0.20 & 0.17 & 30.0$^{(b)}$ & 17 $\pm$ 10 & 16 $\pm$ 4    & $18^{+7}_{-4}$  & $19^{+6}_{-5}$         \\
$2_2-1_1$ & G078.88+00.70  & 3.33  & 8.36  & 22 $\pm$ 6     & 0.07 & 0.05 & 45.2$^{(e)}$ & 16 $\pm$ 4  & 1.7 $\pm$ 0.4 & $14^{+4}_{-2}$  & $2.1^{+0.4}_{-0.4}$    \\
$2_2-1_1$ & G081.87+00.78  & 1.30  & 8.26  & 14.0 $\pm$ 1.8 & 0.10 & 0.09 & 20.9$^{(e)}$ & 19 $\pm$ 10 & 5.2 $\pm$ 1.5 & $20^{+9}_{-5}$  & $6.1^{+1.5}_{-1.4}$    \\
$2_2-1_1$ & G105.41+09.87  & 0.89  & 8.62  & 19 $\pm$ 5     & 0.07 & 0.06 & ...          & 16 $\pm$ 8  & 1.1 $\pm$ 0.3 & $13^{+3}_{-2}$  & $1.5^{+0.3}_{-0.3}$    \\
$5_5-4_4$ & G121.29+00.65  & 0.93  & 8.86  & 13 $\pm$ 3     & ...  & ...  & ...          & ...         & ...           & ...             & ...                    \\
$5_5-4_4$ & G123.06-06.30b & 2.38  & 9.84  & 19 $\pm$ 3     & ...  & ...  & ...          & ...         & ...           & ...             & ...                    \\
$5_5-4_4$ & G123.06-06.30a & 2.82  & 10.15 & 17 $\pm$ 4     & ...  & ...  & ...          & ...         & ...           & ...             & ...                    \\
$5_5-4_4$ & G133.94+01.06  & 1.95  & 9.80  & 11.8 $\pm$ 1.0 & 0.22 & 0.25 & 28.5$^{(i)}$ & 19 $\pm$ 10 & 4.8 $\pm$ 1.2 & $20^{+10}_{-5}$ & $5.6^{+1.5}_{-1.2}$    \\
$5_5-4_4$ & G351.44+00.65  & 1.33  & 7.03  & 23 $\pm$ 3     & ...  & ...  & ...          & ...         & ...           & ...             & ...                    \\
$5_5-4_4$ & G359.61-00.24  & 2.67  & 5.67  & 12.9 $\pm$ 1.8 & ...  & ...  & ...          & ...         & ...           & ...             & ...                    \\
$5_5-4_4$ & G000.67-00.03  & 7.75  & 0.60  & 5.8 $\pm$ 0.3  & ...  & ...  & ...          & ...         & ...           & ...             & ...                    \\
$5_5-4_4$ & G005.88-00.39  & 2.99  & 5.37  & 5.9 $\pm$ 0.2  & 0.23 & 0.26 & 39.7$^{(f)}$ & 18 $\pm$ 9  & 8.5 $\pm$ 2.0 & $19^{+9}_{-5}$  & $10^{+3}_{-2}$         \\
$5_5-4_4$ & G009.62+00.19  & 5.15  & 3.37  & 12.3 $\pm$ 1.2 & 0.09 & 0.10 & 33.7$^{(d)}$ & 15 $\pm$ 3  & 2.3 $\pm$ 0.5 & $15^{+6}_{-3}$  & $2.7^{+0.8}_{-0.7}$    \\
$5_5-4_4$ & G010.47+00.02  & 8.55  & 1.55  & 3.8 $\pm$ 0.8  & ...  & ...  & 32.0$^{(j)}$ & ...         & ...           & ...             & ...                    \\
$5_5-4_4$ & G010.62-00.38  & 4.95  & 3.59  & 13.9 $\pm$ 1.0 & 0.20 & 0.20 & 98.0$^{(c)}$ & 18 $\pm$ 7  & 4.9 $\pm$ 1.1 & $18^{+8}_{-4}$  & $5.8^{+1.6}_{-1.4}$    \\
$5_5-4_4$ & G012.88+00.48  & 2.48  & 5.95  & 6.9 $\pm$ 0.8  & 0.10 & 0.11 & 25.8$^{(h)}$ & 13 $\pm$ 4  & 1.6 $\pm$ 0.3 & $12^{+4}_{-2}$  & $2.0^{+0.6}_{-0.6}$    \\
$5_5-4_4$ & G012.81-00.19  & 2.92  & 5.54  & 16.8 $\pm$ 1.7 & 0.18 & 0.18 & 20.3$^{(h)}$ & 15 $\pm$ 4  & 2.5 $\pm$ 0.5 & $15^{+6}_{-3}$  & $3.0^{+0.9}_{-0.8}$    \\
$5_5-4_4$ & W33            & 2.92  & 5.54  & 23 $\pm$ 4     & ...  & ...  & 20.3$^{(h)}$ & 17 $\pm$ 4  & 2.5 $\pm$ 0.5 & ...             & ...                    \\
$5_5-4_4$ & G012.90-00.26  & 2.53  & 5.91  & 9.2 $\pm$ 1.9  & ...  & ...  & 26.7$^{(h)}$ & ...         & ...           & ...             & ...                    \\
$5_5-4_4$ & G014.33-00.64  & 1.12  & 7.26  & 19 $\pm$ 3     & 0.16 & 0.16 & 25.5$^{(f)}$ & 18 $\pm$ 14 & 2.4 $\pm$ 0.8 & $17^{+7}_{-4}$  & $2.7^{+0.7}_{-0.6}$    \\
$5_5-4_4$ & G015.03-00.67  & 2.00  & 6.43  & 10.2 $\pm$ 1.7 & 0.04 & 0.04 & 32.9$^{(h)}$ & 25 $\pm$ 20 & 0.6 $\pm$ 0.3 & $22^{+13}_{-6}$ & $0.69^{+0.12}_{-0.14}$    \\
$5_5-4_4$ & G024.78+00.08  & 6.67  & 3.61  & 9.7 $\pm$ 1.0  & ...  & ...  & 26.2$^{(f)}$ & ...         & ...           & ...             & ...                    \\
$5_5-4_4$ & G029.95-00.01  & 4.83  & 4.80  & 8.9 $\pm$ 1.1  & 0.08 & 0.07 & 35.8$^{(h)}$ & 19 $\pm$ 12 & 1.5 $\pm$ 0.5 & $16^{+5}_{-3}$  & $1.8^{+0.3}_{-0.3}$    \\
$5_5-4_4$ & G030.70-00.06  & 6.54  & 4.31  & 10 $\pm$ 3     & ...  & ...  & 25.0$^{(j)}$ & ...         & ...           & ...             & ...                    \\
$5_5-4_4$ & G031.24-00.11  & 13.16 & 7.42  & 4.9 $\pm$ 0.5  & ...  & ...  & 41.3$^{(h)}$ & ...         & ...           & ...             & ...                    \\
$5_5-4_4$ & G032.79+00.19  & 9.71  & 5.26  & 8.6 $\pm$ 0.7  & ...  & ...  & 19.7$^{(e)}$ & ...         & ...           & ...             & ...                    \\
$5_5-4_4$ & G032.74-00.07  & 7.94  & 4.60  & 19 $\pm$ 4     & ...  & ...  & 28.5$^{(h)}$ & ...         & ...           & ...             & ...                    \\
$5_5-4_4$ & G034.41+00.23  & 2.94  & 6.14  & 14 $\pm$ 3     & ...  & ...  & 26.5$^{(g)}$ & ...         & ...           & ...             & ...                    \\
$5_5-4_4$ & G034.30+00.15  & 10.58 & 5.70  & 9.6 $\pm$ 0.7  & 0.26 & 0.24 & 36.2$^{(g)}$ & 18 $\pm$ 8  & 5.6 $\pm$ 1.4 & $17^{+5}_{-3}$  & $7.2^{+1.5}_{-1.4}$    \\
$5_5-4_4$ & G035.02+00.34  & 2.33  & 6.57  & 15 $\pm$ 3     & 0.06 & 0.06 & 31.2$^{(h)}$ & 19 $\pm$ 13 & 1.1 $\pm$ 0.4 & $15^{+4}_{-3}$  & $1.3^{+0.3}_{-0.2}$    \\
$5_5-4_4$ & G037.42+01.51  & 1.88  & 6.94  & 9.9 $\pm$ 1.5  & 0.10 & 0.11 & ...          & 13 $\pm$ 9  & 1.3 $\pm$ 0.3 & $12^{+3}_{-2}$  & $1.6^{+0.5}_{-0.5}$    \\
$5_5-4_4$ & G035.19-00.74  & 2.19  & 6.67  & 15 $\pm$ 2     & 0.07 & 0.07 & 24.2$^{(e)}$ & 16 $\pm$ 13 & 1.6 $\pm$ 0.4 & $15^{+5}_{-3}$  & $1.8^{+0.5}_{-0.5}$    \\
$5_5-4_4$ & G035.20-01.73  & 2.43  & 6.51  & 8.3 $\pm$ 1.0  & 0.06 & 0.06 & 20.0$^{(b)}$ & 17 $\pm$ 6  & 1.1 $\pm$ 0.3 & $16^{+6}_{-3}$  & $1.3^{+0.3}_{-0.3}$    \\
$5_5-4_4$ & G043.16+00.01  & 11.11 & 7.60  & 5.18 $\pm$ 0.08  & 0.56 & 0.56 & 35.0$^{(b)}$ & 17 $\pm$ 5  & 26 $\pm$ 6    & $18^{+6}_{-4}$  & $35^{+9}_{-7}$         \\
$5_5-4_4$ & G043.79-00.12  & 6.02  & 5.77  & 5.8 $\pm$ 0.5  & 0.07 & 0.07 & 38.2$^{(h)}$ & 14 $\pm$ 12 & 1.6 $\pm$ 0.5 & $12^{+3}_{-2}$  & $1.9^{+0.5}_{-0.5}$    \\
$5_5-4_4$ & G045.07+00.13  & 7.75  & 6.19  & 9.7 $\pm$ 0.8  & 0.11 & 0.11 & 51.6$^{(h)}$ & 19 $\pm$ 26 & 4 $\pm$ 2     & $16^{+6}_{-3}$  & $4.3^{+1.1}_{-1.0}$    \\
$5_5-4_4$ & G048.60+00.02  & 10.75 & 8.16  & 4.9 $\pm$ 0.4  & 0.05 & 0.05 & 35.4$^{(h)}$ & 16 $\pm$ 9  & 1.2 $\pm$ 0.3 & $13^{+3}_{-2}$  & $1.5^{+0.3}_{-0.3}$    \\
$5_5-4_4$ & G048.99-00.29  & 5.62  & 6.29  & 19 $\pm$ 3     & ...  & ...  & 44.2$^{(h)}$ & ...         & ...           & ...             & ...                    \\
$5_5-4_4$ & G049.48-00.36  & 5.13  & 6.35  & 7.3 $\pm$ 0.4  & 0.22 & 0.22 & 42.9$^{(h)}$ & 14 $\pm$ 5  & 6.9 $\pm$ 1.4 & $14^{+5}_{-3}$  & $9^{+3}_{-2}$          \\
$5_5-4_4$ & G049.49-00.37  & 5.13  & 6.35  & 7.1 $\pm$ 0.4  & ...  & ...  & 42.9$^{(h)}$ & 14 $\pm$ 3  & 6.2 $\pm$ 1.3 & ...             & ...                    \\
$5_5-4_4$ & G049.48-00.38  & 5.41  & 6.34  & 9.5 $\pm$ 0.7  & 0.47 & 0.50 & 30.0$^{(b)}$ & 17 $\pm$ 10 & 16 $\pm$ 4    & $18^{+7}_{-4}$  & $19^{+6}_{-5}$         \\
$5_5-4_4$ & G059.78+00.06  & 2.16  & 7.49  & 16 $\pm$ 5     & 0.13 & 0.14 & 22.1$^{(e)}$ & 11 $\pm$ 3  & 1.2 $\pm$ 0.2 & $10^{+2}_{-2}$  & $1.5^{+0.5}_{-0.5}$    \\
$5_5-4_4$ & G069.54-00.97  & 2.46  & 7.83  & 11.7 $\pm$ 1.8 & 0.07 & 0.07 & 24.7$^{(e)}$ & 17 $\pm$ 9  & 1.1 $\pm$ 0.3 & $15^{+4}_{-3}$  & $1.4^{+0.3}_{-0.3}$    \\
$5_5-4_4$ & G075.29+01.32  & 9.26  & 10.77 & 8.8 $\pm$ 1.5  & 0.03 & 0.03 & ...          & 16 $\pm$ 8  & 0.29 $\pm$ 0.07 & $13^{+3}_{-2}$  & $0.35^{+0.08}_{-0.06}$ \\
$5_5-4_4$ & G075.76+00.33  & 3.51  & 8.21  & 13.1 $\pm$ 1.8 & 0.07 & 0.07 & 45.1$^{(e)}$ & 19 $\pm$ 11 & 1.2 $\pm$ 0.4 & $18^{+7}_{-4}$  & $1.4^{+0.3}_{-0.3}$    \\
$5_5-4_4$ & G075.78+00.34  & 3.83  & 8.28  & 14.1 $\pm$ 1.1 & 0.12 & 0.12 & 45.1$^{(e)}$ & 17 $\pm$ 9  & 2.5 $\pm$ 0.7 & $17^{+6}_{-3}$  & $2.9^{+0.7}_{-0.7}$    \\
$5_5-4_4$ & G078.88+00.70  & 3.33  & 8.36  & 8.9 $\pm$ 0.7  & 0.12 & 0.12 & 45.2$^{(e)}$ & 16 $\pm$ 4  & 1.7 $\pm$ 0.4 & $14^{+4}_{-2}$  & $2.1^{+0.4}_{-0.4}$    \\
$5_5-4_4$ & G079.87+01.17  & 1.61  & 8.21  & 20 $\pm$ 3     & 0.08 & 0.08 & 26.9$^{(e)}$ & 17 $\pm$ 7  & 1.0 $\pm$ 0.2 & $16^{+6}_{-3}$  & $1.1^{+0.3}_{-0.3}$    \\
$5_5-4_4$ & G081.87+00.78  & 1.30  & 8.26  & 9.8 $\pm$ 0.6  & 0.25 & 0.26 & 20.9$^{(e)}$ & 19 $\pm$ 10 & 5.2 $\pm$ 1.5 & $20^{+9}_{-5}$  & $6.1^{+1.5}_{-1.4}$    \\
$5_5-4_4$ & G081.75+00.59  & 1.50  & 8.26  & 12 $\pm$ 2     & 0.09 & 0.09 & 20.9$^{(e)}$ & 12 $\pm$ 4  & 0.80 $\pm$ 0.17 & $11^{+2}_{-2}$  & $1.1^{+0.2}_{-0.2}$    \\
$5_5-4_4$ & G090.92+01.48  & 5.85  & 10.26 & 10.4 $\pm$ 0.9 & ...  & ...  & ...          & ...         & ...           & ...             & ...                    \\
$5_5-4_4$ & G092.67+03.07  & 1.63  & 8.57  & 19.3 $\pm$ 2.0 & 0.11 & 0.10 & 28.9$^{(a)}$ & 14 $\pm$ 13 & 1.8 $\pm$ 0.4 & $13^{+4}_{-3}$  & $2.1^{+0.7}_{-0.6}$    \\
$5_5-4_4$ & G098.03+01.44  & 2.73  & 9.13  & 20 $\pm$ 5     & ...  & ...  & ...          & ...         & ...           & ...             & ...                    \\
$5_5-4_4$ & G105.41+09.87  & 0.89  & 8.62  & 8.6 $\pm$ 0.8  & 0.11 & 0.10 & ...          & 16 $\pm$ 8  & 1.1 $\pm$ 0.3 & $13^{+3}_{-2}$  & $1.5^{+0.3}_{-0.3}$    \\
$5_5-4_4$ & WB89 171       & 1.63  & 8.77  & 26 $\pm$ 6     & ...  & ...  & ...          & 13 $\pm$ 2  & 1.5 $\pm$ 0.3 & ...             & ...                    \\
$5_5-4_4$ & G108.59+00.49  & 2.47  & 9.42  & 10 $\pm$ 4     & 0.04 & 0.04 & 28.4$^{(e)}$ & 18 $\pm$ 6  & 0.52 $\pm$ 0.12 & $16^{+7}_{-3}$  & $0.61^{+0.14}_{-0.14}$    \\
$5_5-4_4$ & G109.87+02.11  & 0.81  & 8.65  & 11.6 $\pm$ 0.9 & 0.14 & 0.14 & 29.9$^{(e)}$ & 20 $\pm$ 6  & 2.4 $\pm$ 0.6 & $19^{+7}_{-4}$  & $3.0^{+0.5}_{-0.5}$    \\
$5_5-4_4$ & G111.54+00.77  & 2.65  & 9.63  & 15 $\pm$ 2     & 0.08 & 0.08 & 30.8$^{(e)}$ & 24 $\pm$ 10 & 1.4 $\pm$ 0.4 & $22^{+10}_{-6}$ & $1.7^{+0.2}_{-0.3}$    \\
$5_5-4_4$ & G111.23-01.23  & 3.33  & 10.04 & 6.6 $\pm$ 0.8  & 0.05 & 0.05 & ...          & 17 $\pm$ 13 & 0.60 $\pm$ 0.19 & $16^{+5}_{-3}$  & $0.69^{+0.17}_{-0.16}$    \\
\enddata
\tablecomments{Column (1): transition, column (2): source name, column (3): the heliocentric distance, column (4): galactocentric distance, column (5): intensity ratio of SO to $^{34}$SO, column (6): the optical depth of SO estimated from LTE method, column(7): the optical depth of SO estimated from non-LTE method, column (8): kinetic temperature, columns (9) and (10): rotational temperature and column density obtained from rotation diagram method assuming optically thin, column (11) and (12): rotational temperature and column density obtained from ``best-fit" model (with optical depth correction, see details in Section \ref{sec:tau})\\
References of kinetic temperatures:
(a) {\citet{Molinari1996}},
 (b) {\citet{Milam2005}},
 (c) {\citet{Hill2010}},
 (d) {\citet{Dunham2011}},
 (e) {\citet{Urquhart2011}},
 (f) {\citet{Wienen2012}},
 (g) {\citet{Cyganowski2013}},
 (h) {\citet{Svoboda2016}},
 (i) {\citet{Keown2019}},
 (j) {\citet{Chen2021}.}}
\end{deluxetable*}
}

For 89 sources with at least two SO line detections, the SO column density ($N_{rot}$) and the rotational temperature ($T_{rot}$) can be derived from the integrated line intensities by performing a standard rotational diagram (Figure \ref{fig:rtd}). For the 7 sources in which both $2_2-1_1$ and $5_5-4_4$ transitions are covered, we can also perform the standard rotational diagram analysis for their $^{34}$SO lines (Figure \ref{fig:rtd34SO}). However, the column densities estimated here assuming optically thin emission should be considered as lower limits in the case where the lines are optically thick. Therefore, following the method described in \citet{Nummelin2000}, we further estimated the optical depth of SO for 67 sources with at least three SO line detections. A brief description of how to estimate the optical depth is provided below.

Assuming that the line has a Gaussian shape, the optical depth at the center of a line, $\tau_{ul}$, can be calculated through
\begin{equation}
\tau_{ul} = \sqrt{\frac{\ln2}{16\pi^3}}\frac{c^3A_{ul}g_uN}{\nu_{ul}^3Q(T_{rot})\Delta V}e^{-E_u/kT_{rot}}(e^{h\nu_{ul}/kT_{rot}}-1),
\label{e:tau}
\end{equation}
where $c$ is the speed of light, $N$ is the total molecular column density, $Q(T_{rot})$ is the molecular partition function evaluated at a temperature $T_{rot}$, $k$ is Boltzmann's constant, $\Delta V$ is the line width. The $Q(T_{rot})$ of SO is taken from JPL Molecular Spectroscopy Catalog\footnote{https://spec.jpl.nasa.gov/ftp/pub/catalog/doc/d048001.pdf} \citep{Pickett1998}: $Q({\rm SO}) = 2.88T_{rot}-15.63$.
Assuming that the sources are homogeneous and the relative distribution of the population over all of the energy levels can be described by a single Boltzmann temperature, the antenna main-beam brightness temperature ($T_{\rm mb}$) can be expressed as
\begin{equation}
T_{\rm mb} = \eta_{\rm bf}[J_\nu(T_{rot})-J_\nu(T_{\rm bg})](1-e^{-\tau_{ul}}),
\label{e:Tmb}
\end{equation}
where
\begin{equation}
J_\nu(T) = \frac{h\nu_{ul}}{k}\frac{1}{e^{h\nu_{ul}/kT}-1},
\label{e:Jnu}
\end{equation}
and where $h$ is Planck's constant. The temperature of the background radiation, $T_{\rm bg}$, was set to 2.73 K. The ``beam-filling factor'', $\eta_{bf}$, could not be directly estimated from the data and was thus set to unity in our calculation. In this case, the results obtained will be beam-averaged. Given the free parameters $N$ and $T_{rot}$, the integrated line intensity can be calculated through equations \eqref{e:tau}-\eqref{e:Jnu}. The set of input parameters producing the closest match to the full set of observed line intensities is called ``best-fit'' model, which is obtained by finding the minimum of the reduced $\chi^2$ function ($\chi_\nu^2$). This is defined as
\begin{equation}
\chi_\nu^2 = \frac{\chi^2}{n-p} = \frac{1}{n-p}\sum\limits_{i=1}^{n}\left(\frac{I^{obs}_i-I^{calc}_i}{\sigma^{obs}_i}\right)^2,
\label{e:chi2}
\end{equation}
where $n$ is the number of data points, $p$ is the number of free parameters, $I^{obs}_i$ is the observed integrated line intensity, $I^{calc}_i$ is the integrated line intensity calculated from equations\eqref{e:tau}-\eqref{e:Jnu}, and $\sigma^{obs}_i$ is the 1 $\sigma$ uncertainty of the observed line intensity (assuming a systematic uncertainty of 20\% for possible errors in pointing and calibration). The intensities from our observations and resulting from the ``best-fit" model are presented in rotation diagram format in Figure \ref{fig:rtd}. Thus we obtained the rotational temperature, the column density, and the optical depth for the above-mentioned 67 sources under LTE assumption. Results for a subsample of 35 sources with measured $^{32}$SO/$^{34}$SO ratios are listed in Table \ref{tab:ratio}, while results for all other sources are presented in Table \ref{tab:dparameter} (denoted as $T_{\rm LTE}$, $N_{\rm LTE}$, to distinguish them from those obtained with standard rotation diagrams assuming optically thin). It is worth to mention that in some sources the observed line intensities for SO $5_6-4_5$ line ($E_u$ = 35 K; with a smaller beam size of $\sim$10$\arcsec$) are higher than that from the ``best-fit" model. This may imply that the beam dilution effect cannot be neglected in these sources. Further mapping observations will be helpful in quantifying the beam-filling factor. As mentioned in Section \ref{sec:d&r}, we favor a single Gaussian fit. While fitting spectra with multiple components might be necessary for sources with non-Gaussian wings (possibly due to an outflow), this may introduces additional uncertainty. As seen in previous work (e.g., \citealt{Chernin1994}) and our results, the wings can be more prominent in higher-transition lines (e.g., G034.30+00.15 in Figure \ref{fig:both2-1} and \ref{fig:both5-4}).

\begin{figure*}[ht]
  \centering
  \includegraphics[width=\textwidth, page=1]{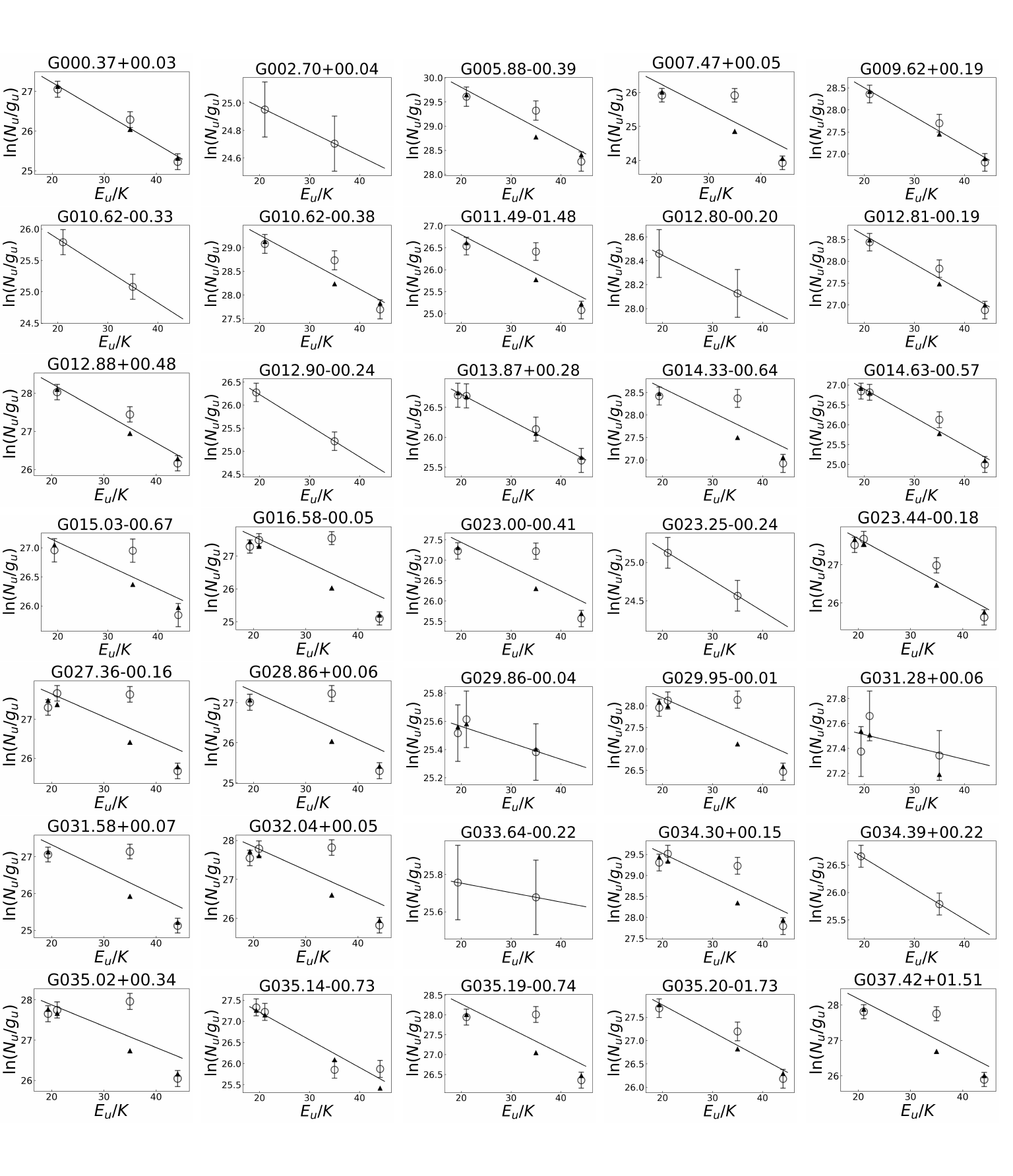}
  \caption{Rotational diagrams of SO for 89 sources with at least two SO line detections. For those 67 sources with at least three SO line detections, the observed line intensities are plotted (circles) together with the intensities resulting from the best-fit model (triangles).}
  \label{fig:rtd}
\end{figure*}
\addtocounter{figure}{-1}
\begin{figure*}[ht]
  \includegraphics[width=\textwidth, page=2]{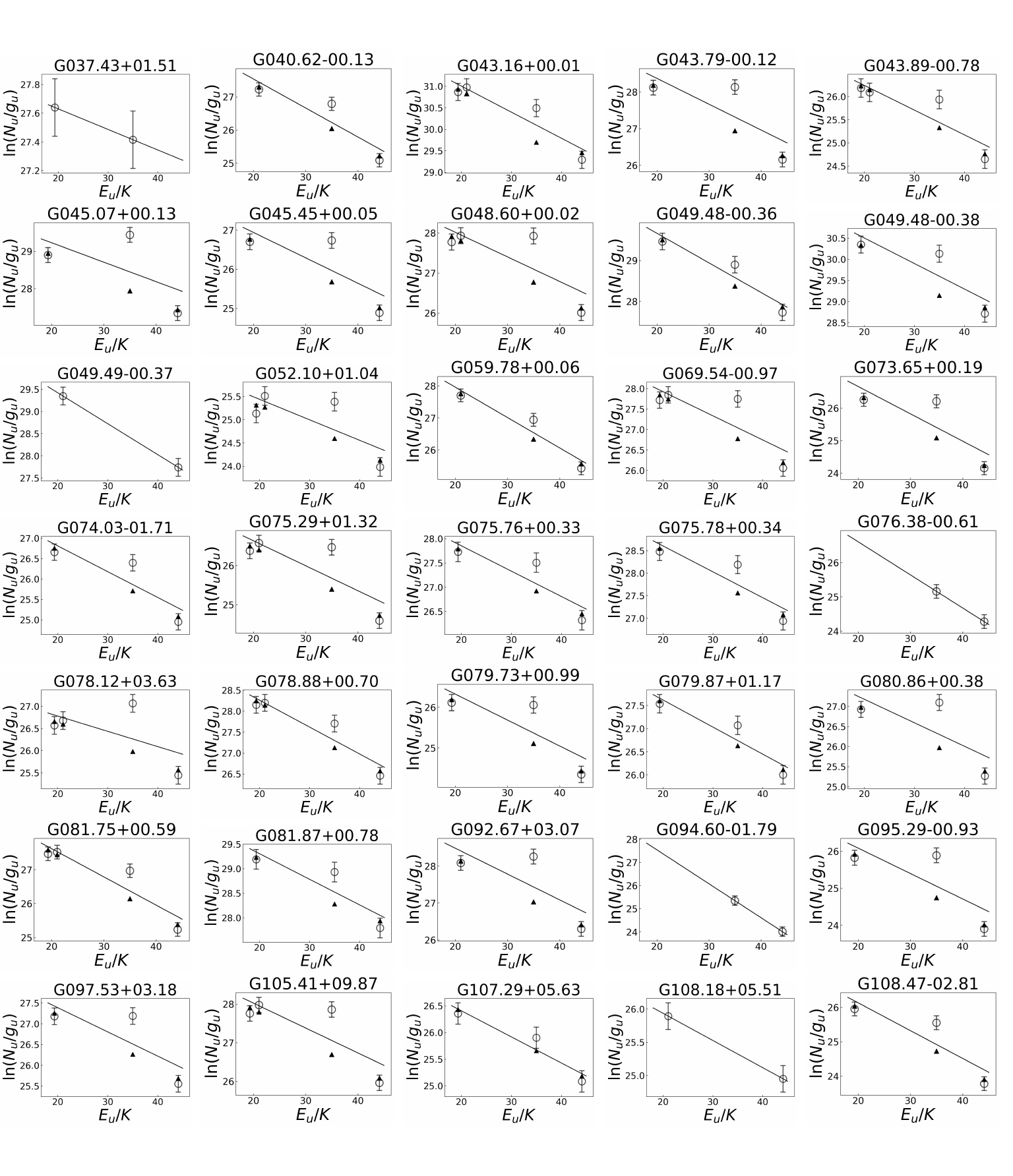}
  \caption{continued.}
\end{figure*}
\addtocounter{figure}{-1}
\begin{figure*}[ht]
  \includegraphics[width=\textwidth, page=1]{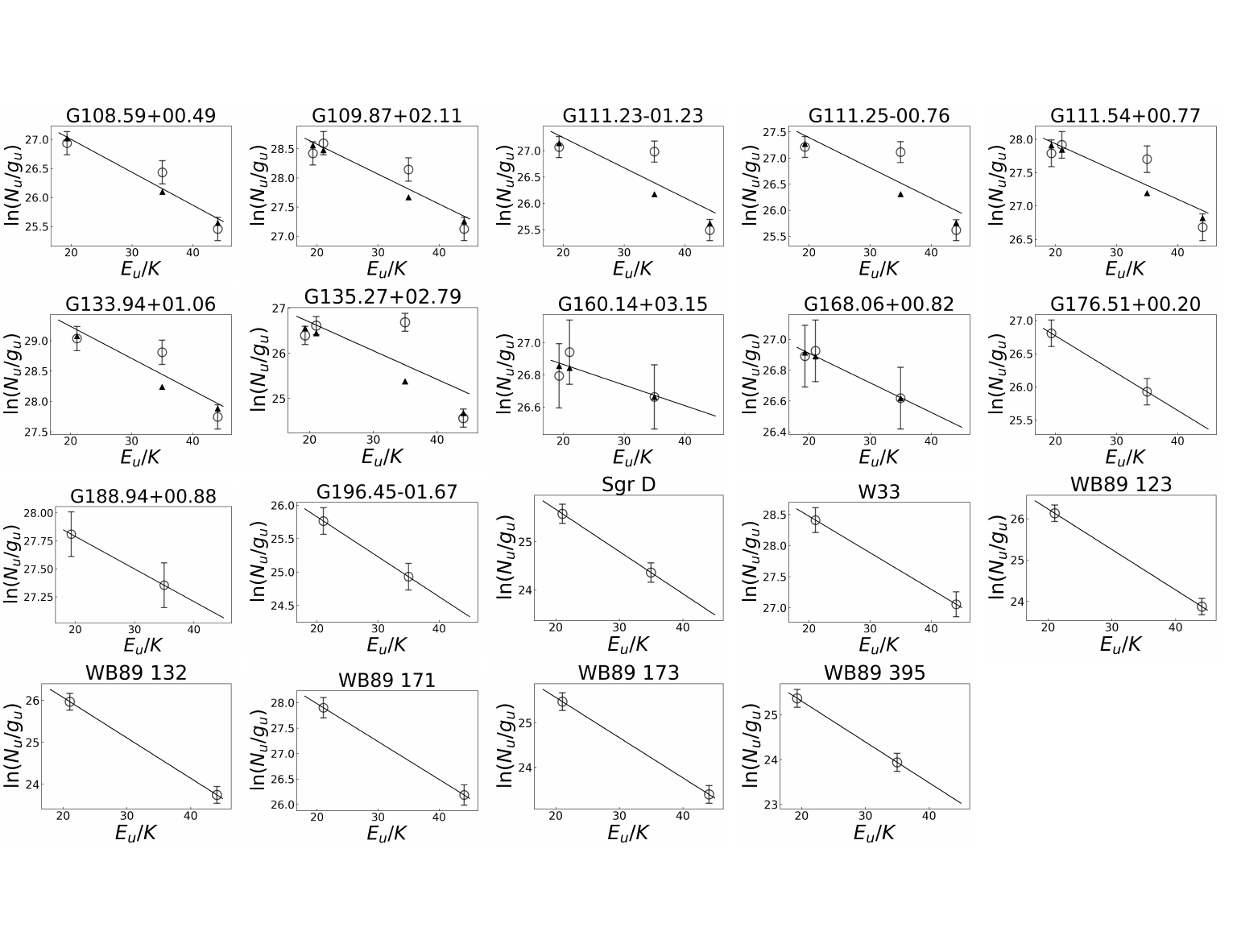}
  \caption{continued.}
\end{figure*}

\begin{figure*}[ht]
  \centering
  \includegraphics[width=\textwidth, page=1]{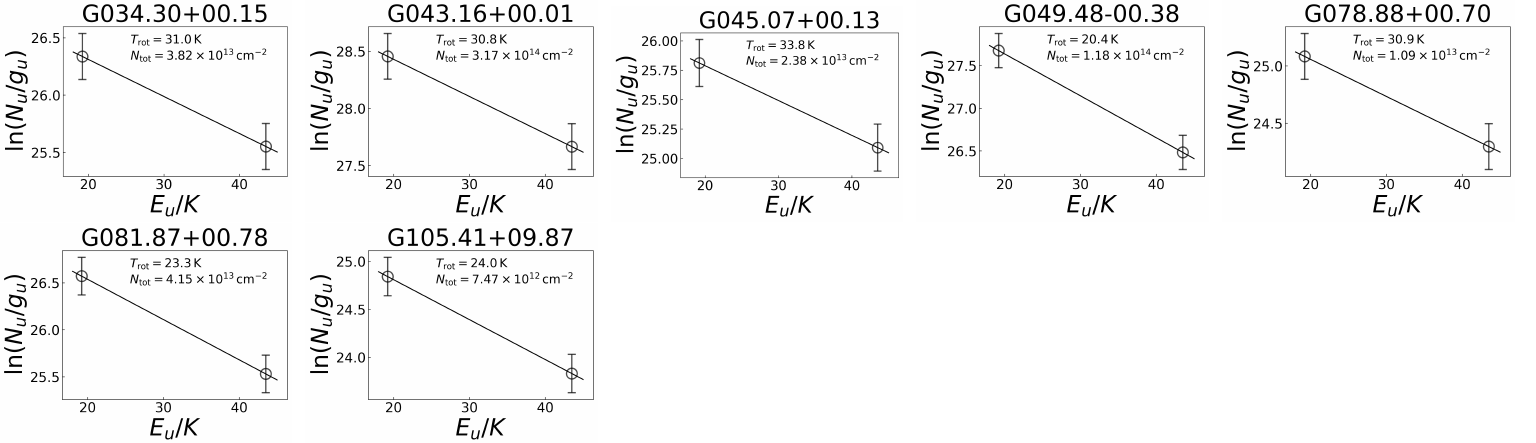}
  \caption{Rotational diagrams of $^{34}$SO for the 7 sources in which both $2_2-1_1$ and $5_5-4_4$ transitions are covered. The $T_{rot}$ and $N_{rot}$ are given directly in each panel.}
  \label{fig:rtd34SO}
  \end{figure*}

\label{figA3}
In some cases, the LTE assumption may not be accurate, e.g., the higher transition lines of SO may not be easily thermalised due to the high critical density (e.g., \citealt{Chernin1994,Wakelam2005}). In addition, the estimation of the optical depth under the LTE assumption will always be a lower limit if optically thin transitions are not covered. Therefore, we also use a non-LTE radiative transfer code (RADEX, \citealt{Van2007}) to further estimate the optical depth. Taking the results from LTE estimate as central values ($T_{\rm LTE}$, $N_{\rm LTE}$), the ranges of kinetic temperature and the SO column density were set to $T_{\rm LTE} \pm 20$ K (or $5 - (T_{\rm LTE}+20)$ K, if $T_{\rm LTE}$ $<$ 25 K) and $10^{\pm 2} N_{\rm LTE}$ cm$^{-2}$, respectively. The range of volume density of H$\rm_2$, $n(\rm H_2)$ was set to $10^{4} - 10^{7}$ cm$^{-3}$. With molecular data from the Leiden Atomic and Molecular Database\footnote{https://home.strw.leidenuniv.nl/~moldata/} (LAMDA) and the line width from our spectral line data, we iterate these three free parameters ($T_k$, $N_{\rm nLTE}$, and $n(\rm H_2)$) to best match (minimise $\chi^2$) the SO line intensities between RADEX and our observations. The estimated non-LTE optical depths for the 35 sources with $^{32}$SO/$^{34}$SO ratio measurements are presented in Table \ref{tab:ratio}.

\begin{figure*}[htpb]
\centering
\includegraphics[width=0.48\textwidth]{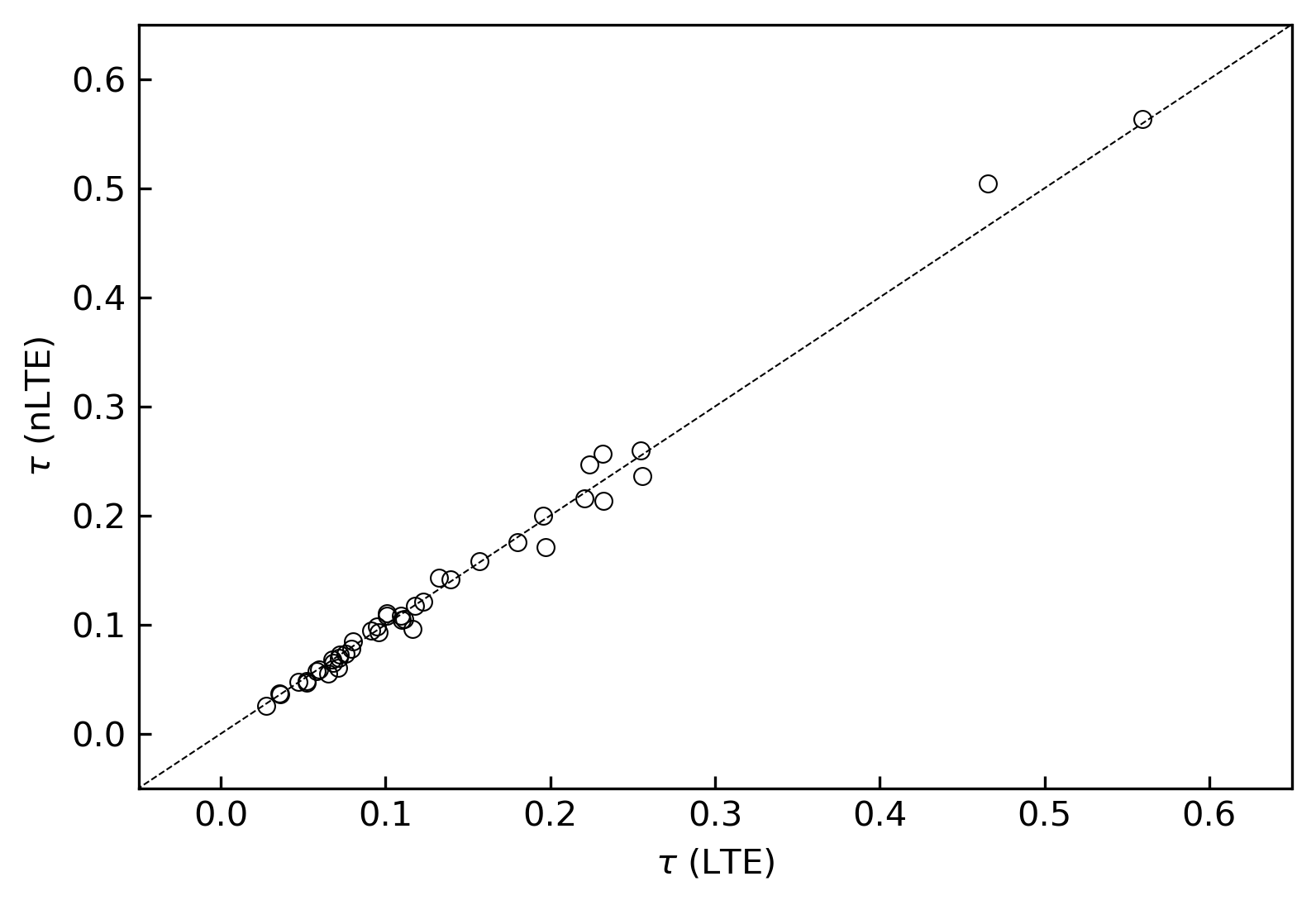}
\caption{The comparison of SO optical depths estimated by LTE and non-LTE methods for the 35 sources with SO/$^{34}$SO ratio measurements.\label{fig:tau_cp}} 
\end{figure*}

Figure \ref{fig:tau_cp} presents a comparison of the optical depths obtained from LTE and non-LTE methods. The optical depths obtained by both methods are consistent, which indicates that deviations from LTE are negligible in our sample. In addition, both methods suggest that most sources ($>$ 90\%) have small optical depth ($<$ 0.3, corresponding correction $<$ 15\% on ratio). This indicates that the optical depth does not have a significant effect on our ratio results.

\subsection{$^{32}$S/$^{34}$S ratios}
\subsubsection{Comparisons of $^{32}$S/$^{34}$S ratios determined from CS and SO species}\label{sec:cp}

The aim of this paper is to explore whether SO species can effectively serve as suitable tracers of the interstellar $^{32}$S/$^{34}$S ratio. Thus, in this section we compare the $^{32}$S/$^{34}$S isotopic ratios inferred from SO measurements and from previous analyses of CS species, and discuss possible explanations of the difference in the results obtained with the two tracers.

\begin{figure*}[htpb]
\centering
\subfigure[]{ \label{fig:cpR1}  
\includegraphics[width=0.48\textwidth]{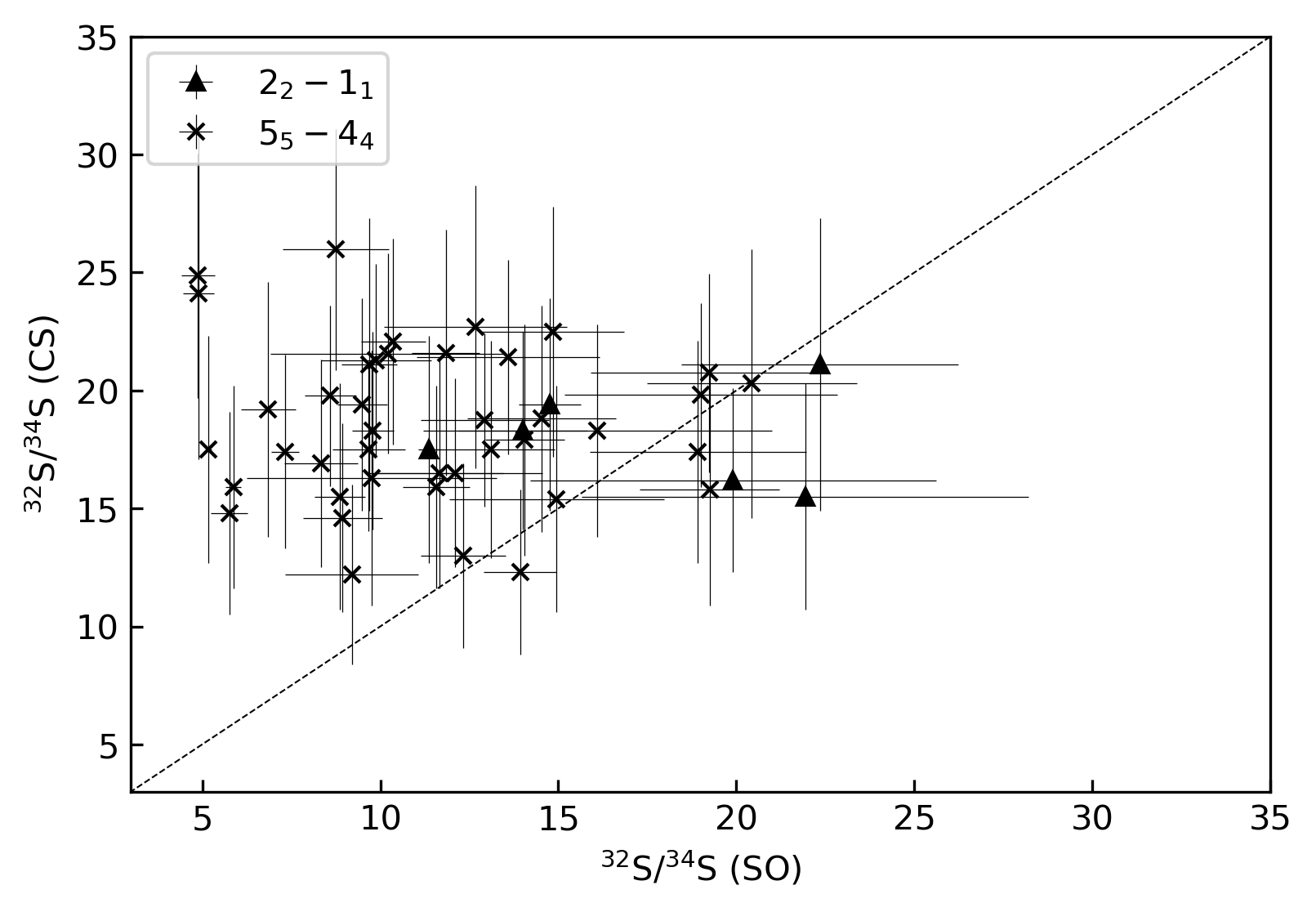}}
\hspace{0.2cm}
\subfigure[]{ \label{fig:cpR2}
\includegraphics[width=0.48\textwidth]{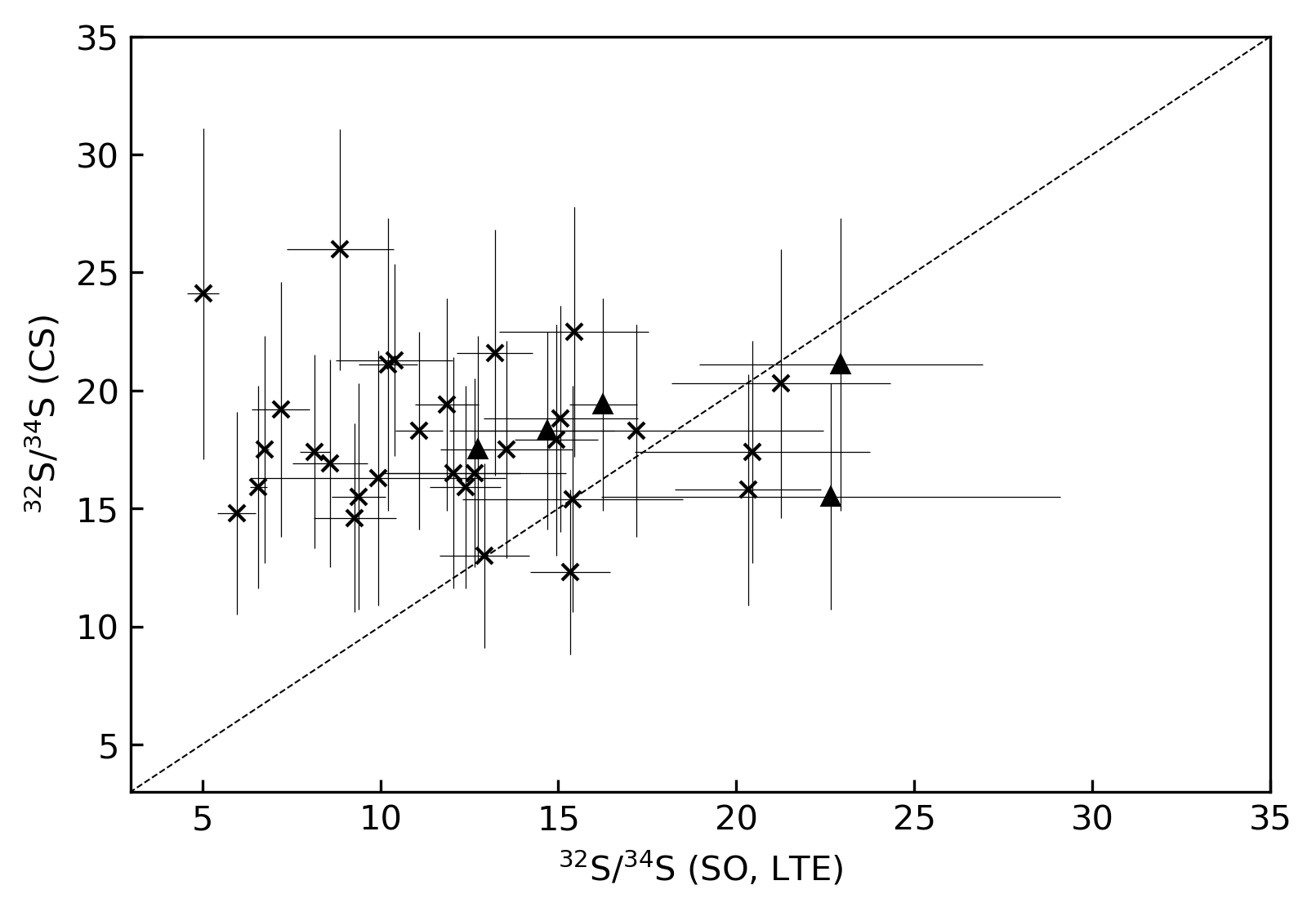}}
\subfigure[]{ \label{fig:cpR3} 
\includegraphics[width=0.48\textwidth]{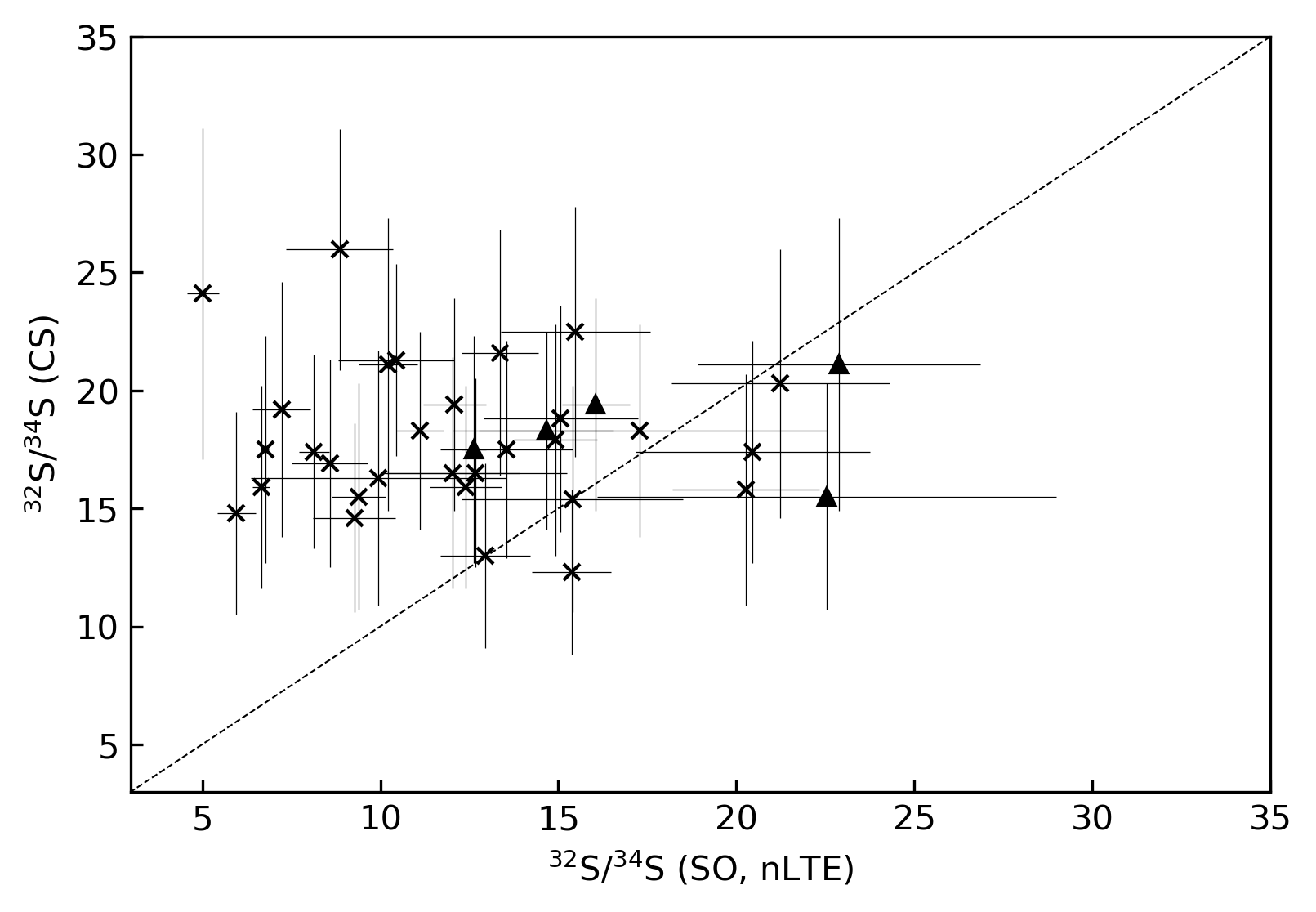}}
\hspace{0.2cm}
\subfigure[]{ \label{fig:cpR4} 
\includegraphics[width=0.48\textwidth]{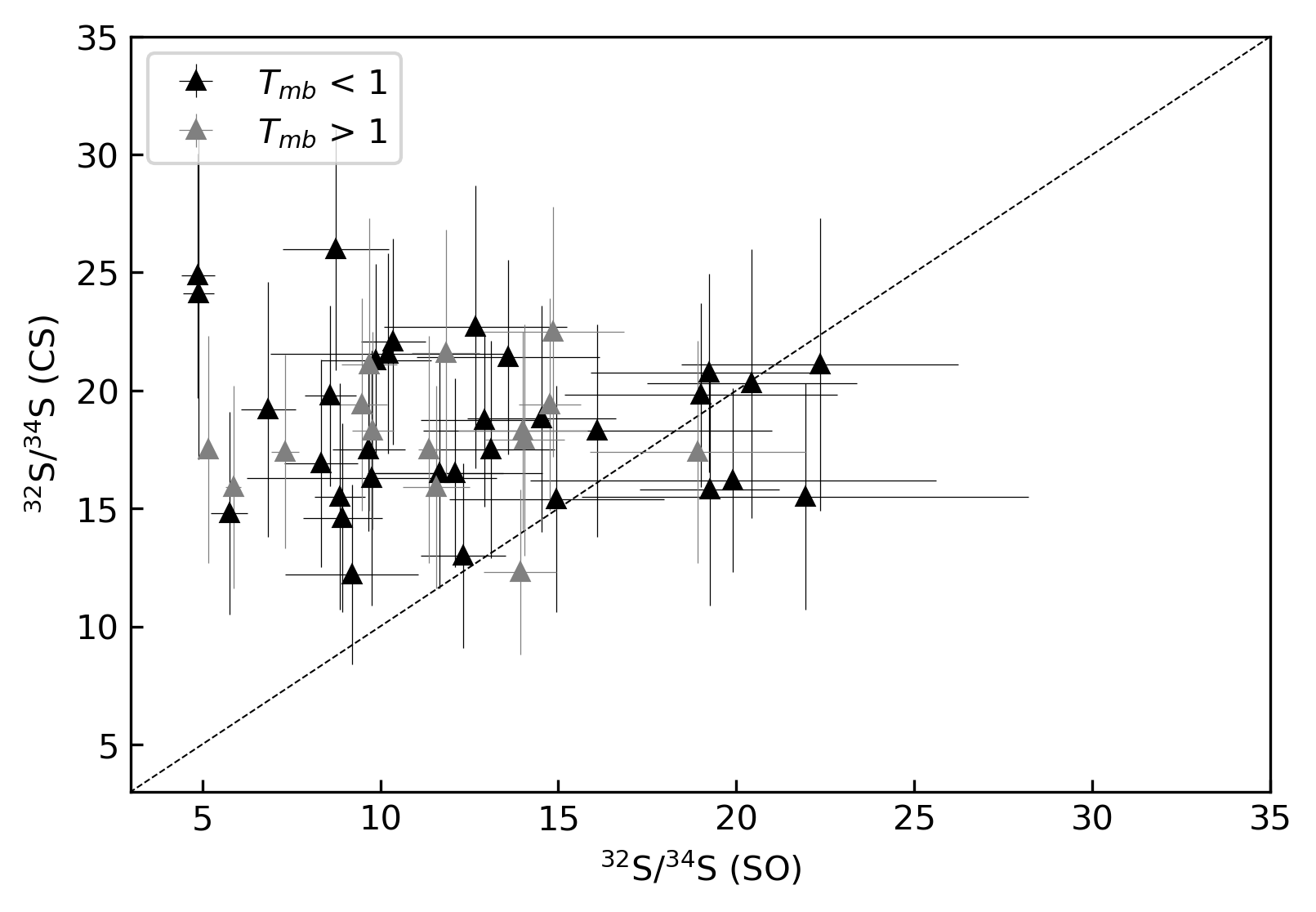}}
\caption{A comparison of $^{32}$S/$^{34}$S ratios from CS ($^{13}$CS/C$^{34}$S$\times$$^{12}$C/$^{13}$C; \citealt{Yu2020, Yan2023}) data and SO ($^{32}$SO/$^{34}$SO; this work). On the $x$ axis, the ratios measured by SO without optical depth corrections are indicated by $^{32}$S/$^{34}$S (SO) (panel(a)), while those with corrections for SO optical depth (see detail in Section \ref{sec:tau}) are indicated by $^{32}$S/$^{34}$S (SO, LTE) (panel(b)), and $^{32}$S/$^{34}$S (SO, nLTE) (panel(c)), respectively. In panel (d), black and grey triangles are sources with SO peak $T_{mb}$ lower and higher than 1 K, respectively. The black dashed lines indicate $y = x$ to guide the eye.}\label{fig:cpR1-4} 
\end{figure*}

Among our targets, 42 have $^{32}$S/$^{34}$S ratios measured from $^{13}$CS and C$^{34}$S (J=2-1 , \citeauthor{Yu2020} \citeyear{Yu2020} and \citeauthor{Yan2023} \citeyear{Yan2023}). In Figure \ref{fig:cpR1}, their ratios are compared with our results from SO data. We find that the ratios measured from CS data are generally higher than those measured from our SO $5_5-4_4$ data. Moreover, the discrepancy still exists after optical depth corrections are applied to 30 out of 42 sources, using both LTE and non-LTE methods, as shown in Figure \ref{fig:cpR2} and \ref{fig:cpR3}. However, the discrepancy is not obvious for our SO $2_2-1_1$ measurements. Reflecting the fact that sources with optically thick lines are expected to exhibit higher peak $T_{mb}$ values, the sources are divided into two sub-samples based on whether they have $T_{mb}$ higher or lower than 1 K, as shown in Figure \ref{fig:cpR4}. We find that there is no clear correlation between different $T_{mb}$ values and the $^{32}$S/$^{34}$S ratios. All of this suggests that the optical depth has an insignificant effect on the ratio difference between SO and CS measurements. However, as mentioned earlier, in cases where optically thin transitions are not covered, the estimated optical depth may represent only a lower limit to the real optical thickness of the lines. Nevertheless, in order to minimize the influence of optical depths when analyzing other factors, the ratios with optical depth correction are used in the following.

\begin{figure*}[htpb]
\centering
\subfigure[]{ \label{fig:cpR5} 
\includegraphics[width=0.48\textwidth]{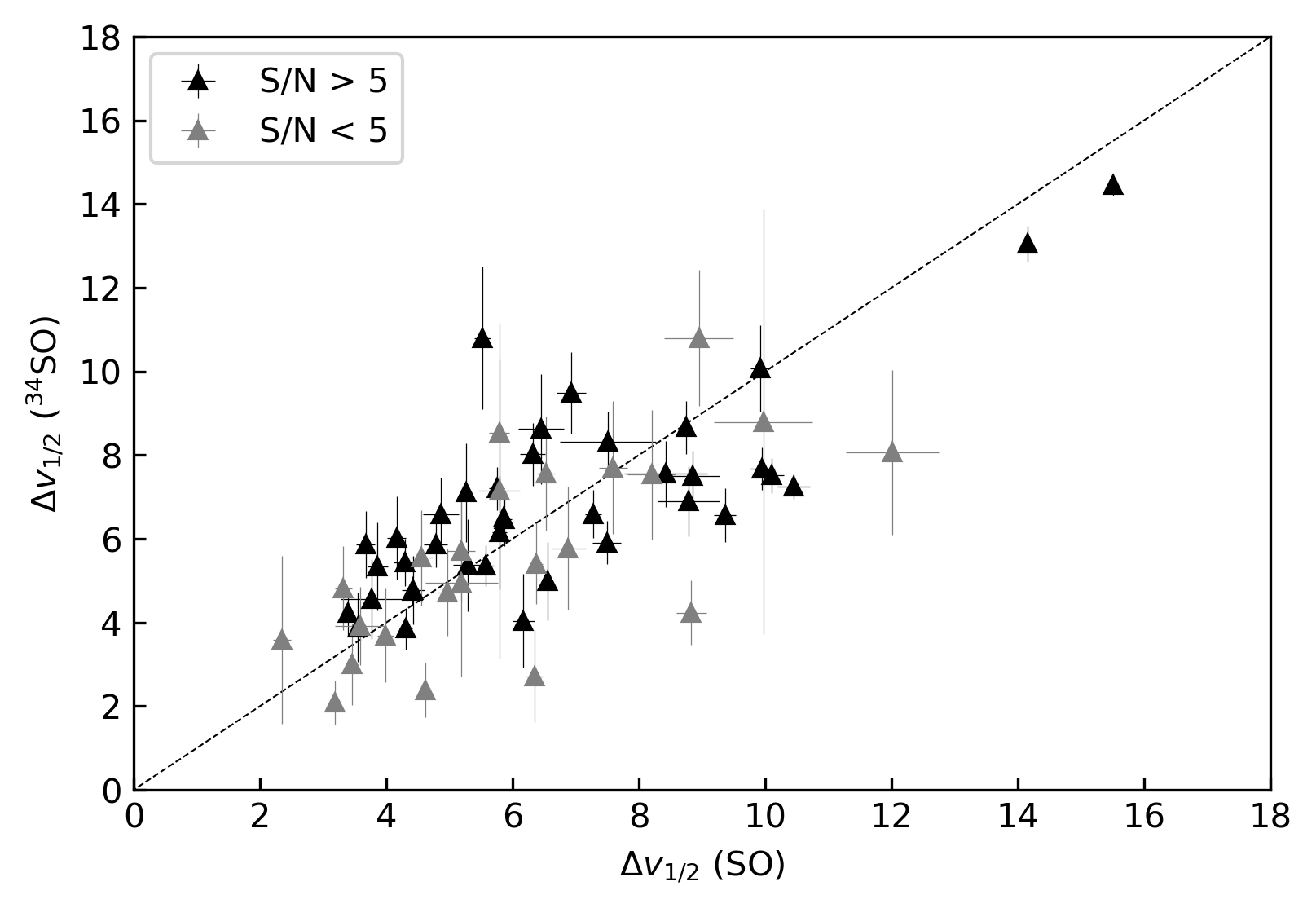}} 
\hspace{0.2cm}
\subfigure[]{ \label{fig:cpR6} 
\includegraphics[width=0.48\textwidth]{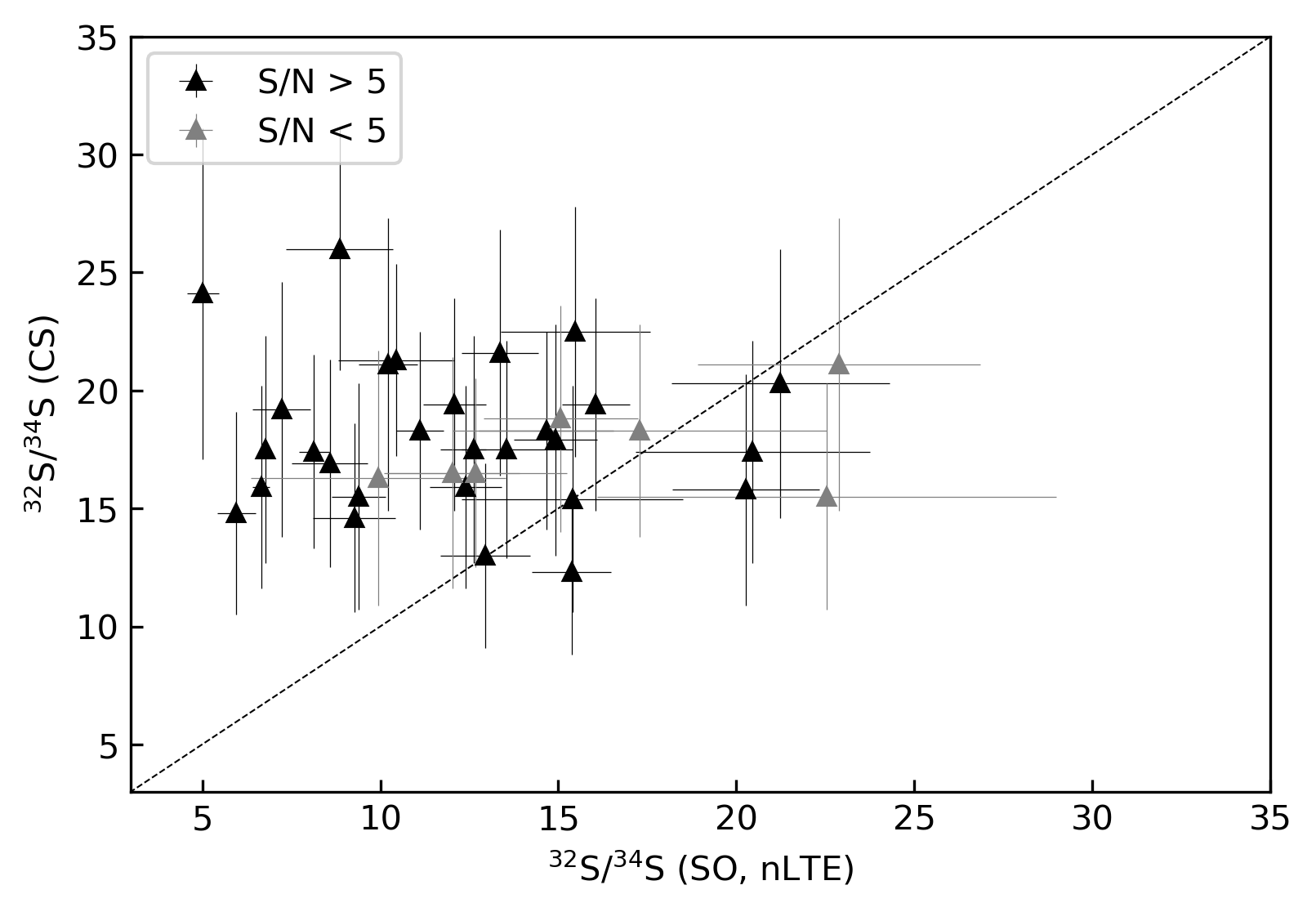}}
\caption{Effect of data quality on the line width and the sulfur isotopic ratio. Panel (a) presents a comparison of line widths of SO and $^{34}$SO, for different data quality. Panel (b) shows a comparison of $^{32}$S/$^{34}$S ratios from CS and SO data for samples with different data quality. Black and grey triangles indicate sources with SNR ($^{34}$SO) higher and lower than 5, respectively. The $y = x$ function is drawn in both panels to guide the eye (black dashed lines).}
\label{fig:cpR5&6}
\end{figure*}

In some cases, the line width and, consequently, the integrated intensity, may be overestimated due to the low quality $^{34}$SO spectrum. In this case, the $^{32}$S/$^{34}$S ratio could be underestimated. In Figure \ref{fig:cpR5&6}, our sources are divided into two sub-samples based on the signal-to-noise ratios of the $^{34}$SO line. A comparison of line widths between SO and $^{34}$SO is presented in Figure \ref{fig:cpR5}. Overall, the line widths of SO and $^{34}$SO are comparable in both samples, indicating no signs of overestimation of the $^{34}$SO line width in the lower quality $^{34}$SO spectra (SNR $<$ 5). Figure \ref{fig:cpR6} compares the measured $^{32}$S/$^{34}$S ratios from SO and CS data for these two samples. The ratios measured by CS species are still systematically higher in the sample with higher signal-to-noise ratios (SNR $>$ 5). This suggests that the data quality is unlikely to be a major factor responsible for the discrepancy of $^{32}$S/$^{34}$S ratios measured from SO and CS lines.

\begin{figure*}[htpb]
\centering
\subfigure[]{ \label{fig:cpR7} 
\includegraphics[width=0.48\textwidth]{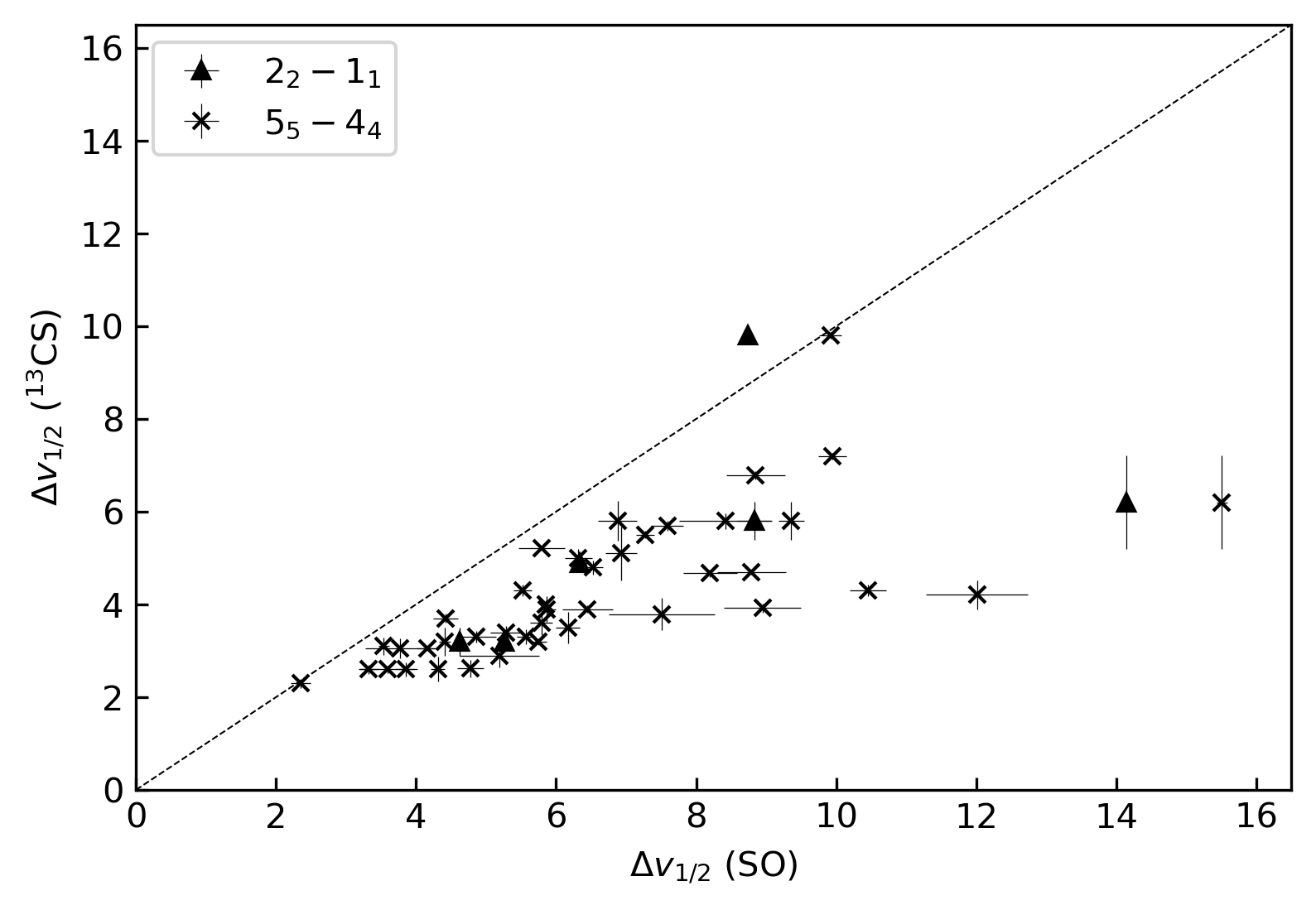}} 
\hspace{0.2cm}
\subfigure[]{ \label{fig:cpR8}  
\includegraphics[width=0.48\textwidth]{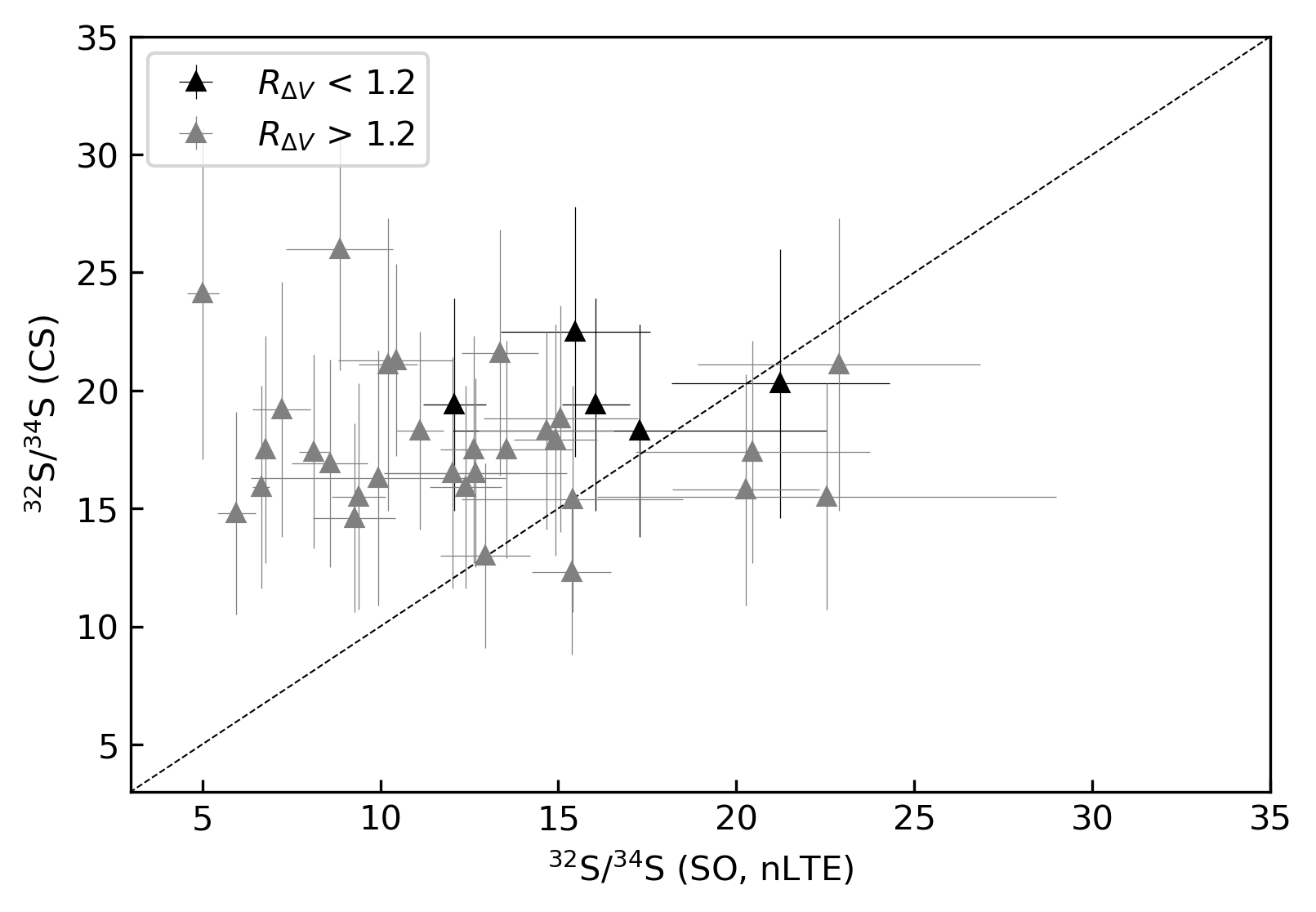}}
\caption{Comparison of the line widths of SO and $^{13}$CS (a) and effect of the line width on the ratio measured from CS and SO data (b). Black triangles are sources with $R_{\Delta V} <$ 1.2 (see text), while grey triangles identify those with $R_{\Delta V}$ $>$ 1.2. The $y = x$ function is drawn in both panels to guide the eye.}\label{fig:cpR7&8} 
\end{figure*}

SO is frequently detected as a tracer of shocked gas associated with the bipolar outflows from protostars \citep{Booth2018, Watt1986, Welch1988, Bachiller2001, Esplugues2014}, while CS and its isotopologues are commonly used to trace dense gas. It is also found that SO and CS have different spatial distributions within molecular clouds (e.g., \citealt{Swade1989, Chernin1994,Nilsson2000}). Therefore, the differences in the $^{32}$S/$^{34}$S ratios measured from the two species may be caused by the different gas components they trace. To assess whether this is the case, we compare the line widths of the SO and $^{13}$CS lines in Figure \ref{fig:cpR7}. The $^{13}$CS J=2-1 line widths are narrower than those of the SO lines ($2_2-1_1$ and $5_5-4_4$). The broader line width of SO indicates its emission may be related to more complex and active gas components within molecular clouds. Namely, besides the relatively quiescent component traced by $^{13}$CS, the SO emission may also include the contribution from a warmer, denser, and more turbulent component (e.g., \citealt{Fontani2023}). In Figure \ref{fig:cpR8}, our sources are divided into two sub-samples according to their SO-to-$^{13}$CS line width ratio ($R_{\Delta V}$). Sources with similar SO and $^{13}$CS line widths ($R_{\Delta V} < $ 1.2, indicating a low contribution from turbulent components) have $^{32}$S/$^{34}$S ratios measured from the two molecular species that do agree each other (though there are only 5 sources). This may indicate that in turbulent (warmer and denser) regions the $^{32}$S/$^{34}$S ratio is lower than in relatively quiescent regions, based on the lower $^{32}$S/$^{34}$S ratios measured from SO lines with respect to the ones obtained from the CS lines. In denser and more turbulent regions, the SO lines could be optically thick, so that the $^{32}$S/$^{34}$S ratio will be underestimated if the optical depth effect is not properly corrected. In warmer regions, the $^{34}$SO line is likely contaminated by COMs. For example, the $^{34}$SO $2_2-1_1$ line could be contaminated by nearby lines of CH$_3$COOH and CH$_3$COCH$_3$ detected in the hot core \citep{Fontani2024}. This could lead to an overestimate of its intensity and, thus, to a lower $^{32}$SO/$^{34}$SO ratio. Indeed, \cite{Artur2023} proposed that the SO/$^{34}$SO ratio might be a good tracer of the inner high-density envelope, and a low value seems to be related to the detection of multiple COMs.

To sum up, the discrepancy between $^{32}$S/$^{34}$S ratios measured from SO and CS isotopologues may be caused by the different gas components traced by the two tracers. But the situation could be more complex than discussed above. By performing two-phase chemical model calculations, \cite{Esplugues2014} found that SO and SO$_2$ reactions can change from being dominated by O$_2$ to being dominated by OH in the shocked gas, leading to a rapid decrease in the SO/SO$_2$ ratio with time. In addition, the line broadening mechanisms can also be complex. \cite{Ginsburg2024} reported a unique millimeter ultra-broad-line object without strong shocks (no detection of SiO), which exhibits a wide line width in SO, CS lines. A recent magnetohydrodynamical simulation suggested that the observed line profiles, including the line width, can be affected by the mean magnetic field direction, while turbulence-induced density variations should be considered when trying to derive chemical abundances from observations \citep{Beitia2024}. However, detailed discussion about the effects of magnetic field, turbulence and shock is beyond the scope of this paper. Further observations and modeling work will be necessary for a more comprehensive understanding.

\subsubsection{$^{32}$S/$^{34}$S ratios across the Galaxy}
\begin{figure*}[htpb]
\centering
\includegraphics[width=0.98\textwidth]{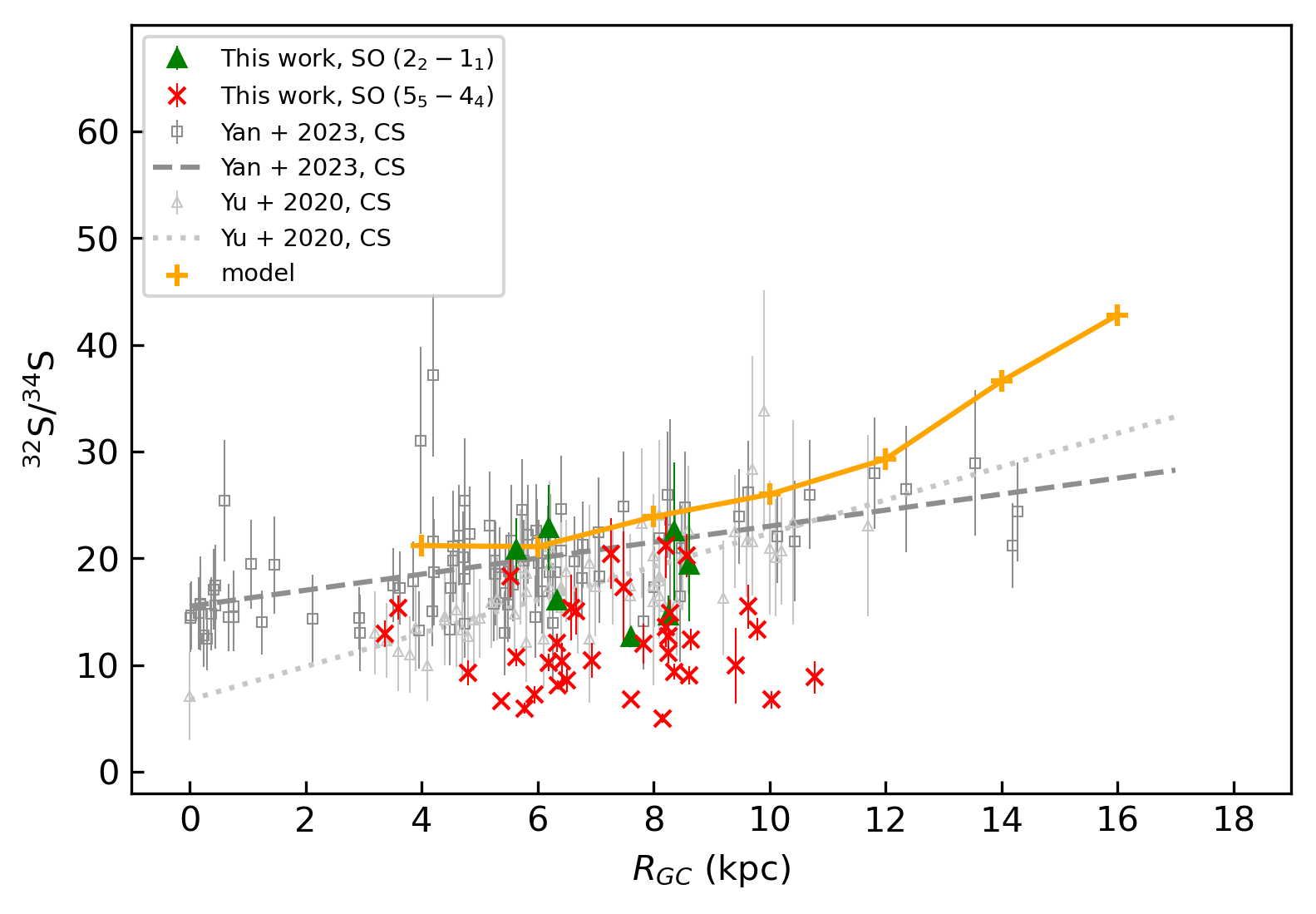}
\caption{$^{32}$S/$^{34}$S isotope ratios as functions of the Galactocentric distance. The filled green triangles and red crosses are our results from measurements of SO/$^{34}$SO $2_2-1_1$ and $5_5-4_4$ lines, respectively. The open light gray triangles and open dark gray squares are values determined from CS species (J=2-1 lines) with the double isotope method, taken from \citet{Yu2020} and \cite{Yan2023}, respectively. The first-order polynomial fits to their data are also plotted as light gray dotted and dark gray dashed lines, respectively. The orange solid curve indicates the predictions of the GCE model of Romano et al. (2025, to be submitted). \label{fig:ratio_rgc}}
\end{figure*}

With the IRAM 30 m and the SMT 10 m measurements of SO and $^{34}$SO $2_2-1_1$ and $5_5-4_4$ lines, we obtained $^{32}$S/$^{34}$S ratios for 56 sources, including 7 sources with ratios measured from both transitions. Among them, 35 sources have optical depth estimates (see details in Table \ref{tab:ratio}); optical depth effects do not affect the estimated ratios significantly. Optical-depth corrected ratios are plotted against their Galactocentric distances in Figure \ref{fig:ratio_rgc}. Ratios measured from CS species taken from \citet{Yu2020} and \cite{Yan2023} are also shown in Figure \ref{fig:ratio_rgc}. Our ratios measured from SO $2_2-1_1$ transition lines (green triangles in Figure \ref{fig:ratio_rgc}) are generally close to previous results obtained from CS species (grey squares and triangles), though the sample is limited. On the other hand, ratios obtained from SO $5_5-4_4$ lines (red crosses) are systematically lower than previous CS measurements. For the 7 sources in which both transitions are covered, the SO $2_2-1_1$ line is optically thinner than the $5_5-4_4$ line. In addition, the $^{34}$SO column density of these sources can be derived from rotational diagrams (see Figure \ref{fig:rtd34SO}). We found that the column density ratios are consistent with the opacity corrected line intensity ratios of $2_2-1_1$ lines within the error range ($\sim$20\%). Thus we propose that the lower transition lines of SO and $^{34}$SO are more suitable for measuring the $^{32}$S/$^{34}$S ratio than the higher transition lines. However, the current sample is limited, and supplementary SO data for sources in the Galactic center region and in the outer Galaxy are badly required to accurately determine the Galactic radial $^{32}$S/$^{34}$S gradient.

Comparing observations with Galactic chemical evolution model predictions can help us to better understand the Galactic radial variation of $^{32}$S/$^{34}$S. The theoretical $^{32}$S/$^{34}$S gradient predicted by the GCE model of Romano et al. (2025, to be submitted) is shown in Figure \ref{fig:ratio_rgc}. In this model, the adopted stellar yields for massive stars are taken from \cite{Limongi2018}, who computed homogeneous stellar yields for a wide range of initial metallicities ([Fe/H] = -3, -2, -1, 0) and initial rotational velocities of the stars ($v_{\rm rot}$~= 0, 150, and 300 km~s$^{-1}$). For the super-solar metallicity regime, the yields are from the homogeneous grids recently provided by the same authors \citep{Roberti2024}. In the 2--12 kpc Galactocentric distance range, the slope of the theoretical gradient agrees very well with that inferred from observations of CS transitions \citep{Yan2023}, though the predicted $^{32}$S/$^{34}$S ratios tend to reproduce the upper envelope of the observations, rather than the average values computed for each Galactocentric distance bin. In the outer Galaxy ($R_{\mathrm{GC}} >$ 12 kpc), the predicted $^{32}$S/$^{34}$S gradient slope steepens, deviating significantly from the observed one, though we note that only a few (no) data exist for $R_{\mathrm GC} >$ 12 (14) kpc.

We know from stellar evolution and nucleosynthesis theory that $^{32}$S originates from both incomplete explosive silicon burning and oxygen burning in massive stars, in proportions that vary from star to star \citep{woosley1995,nomoto2013,Limongi2018}. This makes it a primary element with little dependence on a star's metallicity. On the other hand, $^{34}$S has both a primary and a secondary component. The first one comes from explosive oxygen burning and dominates at low metallicities. The second one originates from the $^{33}$S($n,\gamma$)$^{34}$S reaction and becomes increasingly important at higher metallicities (Marco Limongi, private communication). The different origins of the two isotopes and the dependence of the $^{34}$S yields on metallicity explains why the GCE model predicts lower $^{32}$S/$^{34}$S ratios towards the Galactic center, where the metallicity is higher (an increasing trend in the predicted ratio from 6 to 16 kpc). However, sulfur production in the low-metallicity regime that characterizes the outer Galactic regions is still poorly constrained, which calls for larger datasets for objects found in the outermost regions of the Milky Way.

\section{Summary}\label{sec:sum}
In order to explore whether sulfur monoxide and its isotopologue $^{34}$SO can be suitable tracers of the $^{32}$S/$^{34}$S isotopic ratio, we present the first systematic observations of SO and $^{34}$SO lines toward a large sample of Galactic molecular clouds with accurate distances. Using the IRAM 30 m, 59 (8) out of 82 sources are detected in the SO ($^{34}$SO) $2_2-1_1$ line. With the SMT 10 m, 136 of 184 sources are detected in the SO $5_5-4_4$ line and 55 of the strongest 77 sources are detected in the $^{34}$SO $5_5-4_4$ line. Our main results can be summarized as follows:

\begin{enumerate}
\item From the measured $5_5-4_4$ and $2_2-1_1$ line intensities, $^{32}$S/$^{34}$S ratios are obtained for 56 sources, including 7 sources measured in both transitions. The two telescope beam sizes for these transitions are similar and no systematic variation of the measured isotope ratios is found with increasing the heliocentric distance of the targets; both of these suggest that any observational bias caused by beam dilution effects is negligible. No correlation is found between the measured SO/$^{34}$SO ratio and $T_k$, indicating that fractionation effects is not obvious. Finally, we estimate the optical depths of the SO lines under the assumption of LTE and non-LTE. Both methods consistently suggest that the SO line has a small optical depth and, thus, optical depth effects are unlikely to affect our measured ratios.
\item Overall, our measured $^{32}$S/$^{34}$S ratios from SO and $^{34}$SO lines tend to be systematically lower than those previously estimated from CS data. The discrepancy is all the way present in the sub-sample of high-quality data (SNR $>$ 5) after applying optical depth corrections. But it should be noted that even if the ratios from SO/$^{34}$SO are opacity corrected, they can still be lower limit if optically thin transitions are not covered. The SO lines (with higher upper energy level) present line widths that are systematically wider than $^{13}$CS, indicating that the SO emission may partly come from a warmer, denser, and more turbulent gaseous component. The line widths of $^{13}$CS and SO being equal, the derived ratios are the same. This suggests that the $^{32}$S/$^{34}$S ratios derived from SO lines may be affected by environmental and evolutionary effects. In fact, in warmer regions $^{34}$SO lines are more likely contaminated by COMs, while SO lines can be optically thick. However, these effects can not be quantified with the current dataset. Further investigations, both on the observational and theoretical fronts, are badly needed.
\item Sulfur isotopic ratios measured from the SO $2_2-1_1$ transition are closer to previous results obtained from CS data. For the 7 sources in which both transitions are covered, the SO $2_2-1_1$ line is optically thinner than the $5_5-4_4$ line. Moreover, the column density ratios are consistent with the opacity corrected line intensity ratios of $2_2-1_1$ lines within the error range. Therefore, we propose that the lower transition lines of SO and $^{34}$SO are better suited to measure the $^{32}$S/$^{34}$S ratio than the higher transition lines. However, supplementary SO data for sources lying in both the innermost and the outermost Galactic disc are required to determine the Galactic radial $^{32}$S/$^{34}$S gradient. Comparisons between observations and GCE model predictions suggest that the nucleosynthesis prescriptions need to be revised in the low-metallicity regime, while a satisfactory agreement is found for regions characterized by solar and super-solar metallicities.
\end{enumerate}

\section*{ACKNOWLEDGMENTS}
We thank the operators and staff at both the IRAM 30 m and the SMT 10 m stations for their assistance during our observations. We would like to thank Marco Limongi for his explanation about $^{34}$S synthesis in massive stars. This work is supported by the Natural Science Foundation of China (No. 12373021, 12041302) and the National Key R\&D program of China (2022YFA1603102). Y. P. Z. would like to thank the China Scholarship Council (CSC) for its funding. H. Z. Y. acknowledges the support from the Ministry of Science and Higher Education of the Russian Federation (state contract FEUZ-2023-0019). Y. T. Y. and Y. X. W. are members of the International Max Planck Research School (IMPRS) for Astronomy and Astrophysics at the Universities of Bonn and Cologne. Finally, we wish to thank the anonymous referee for comments that improved the presentation of the paper.
\bibliography{SO_references}{}
\bibliographystyle{aasjounal}

\begin{appendix}

\section{Figures}\label{sec:appendF}
\setcounter{figure}{0}
\renewcommand{\thefigure}{A.\arabic{figure}}
\renewcommand*{\theHfigure}{\thefigure}

\begin{figure*}[ht]
\centering
\includegraphics[width=0.85\textwidth, page=1]{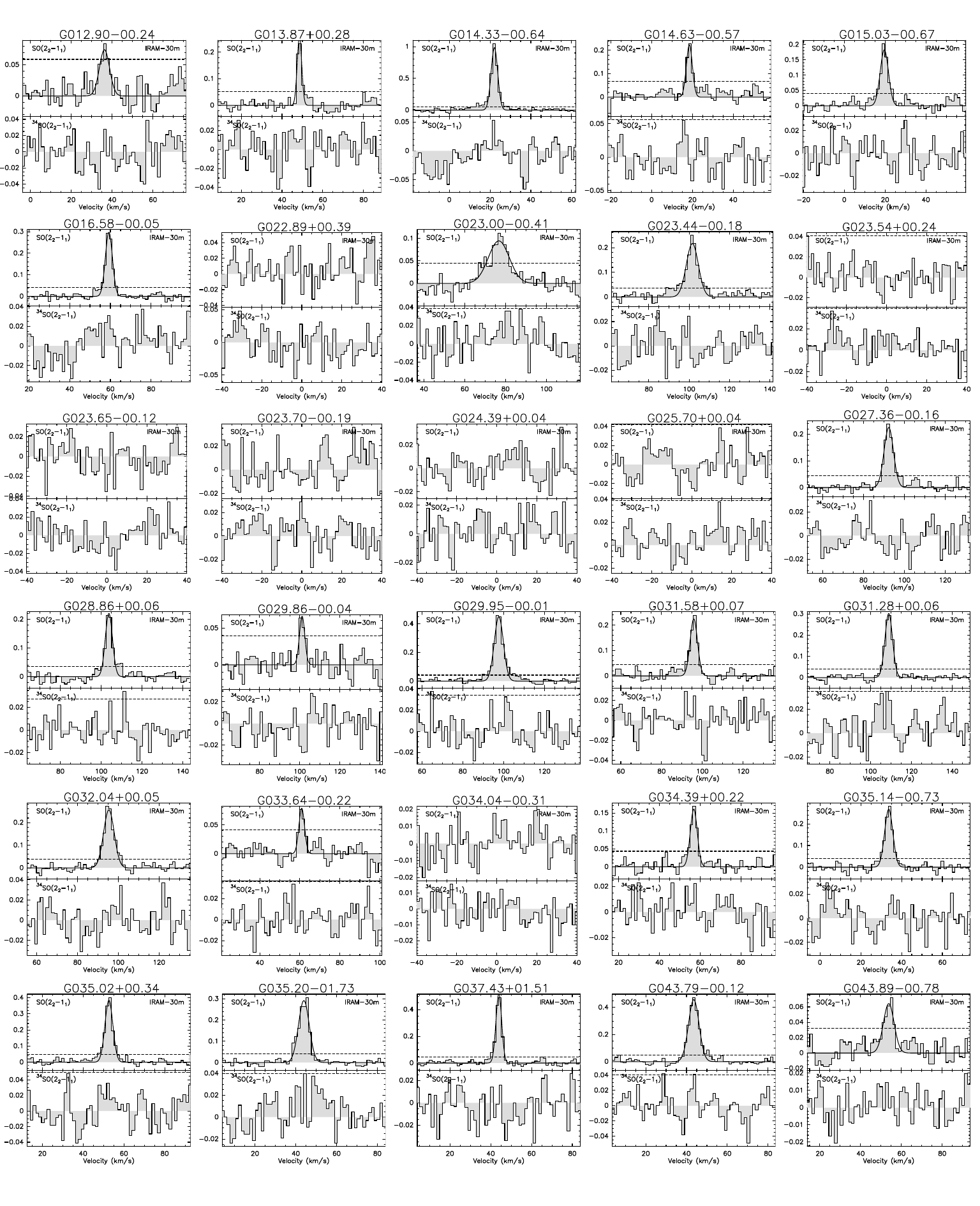}
\caption{The IRAM 30 m spectra of the sources without effective measurements of SO/$^{34}$SO ($2_2-1_1$ lines).}
\label{fig:otherIRAM}
\end{figure*}

\addtocounter{figure}{-1}
\begin{figure*}[ht]
\centering
\includegraphics[width=0.85\textwidth, page=2]{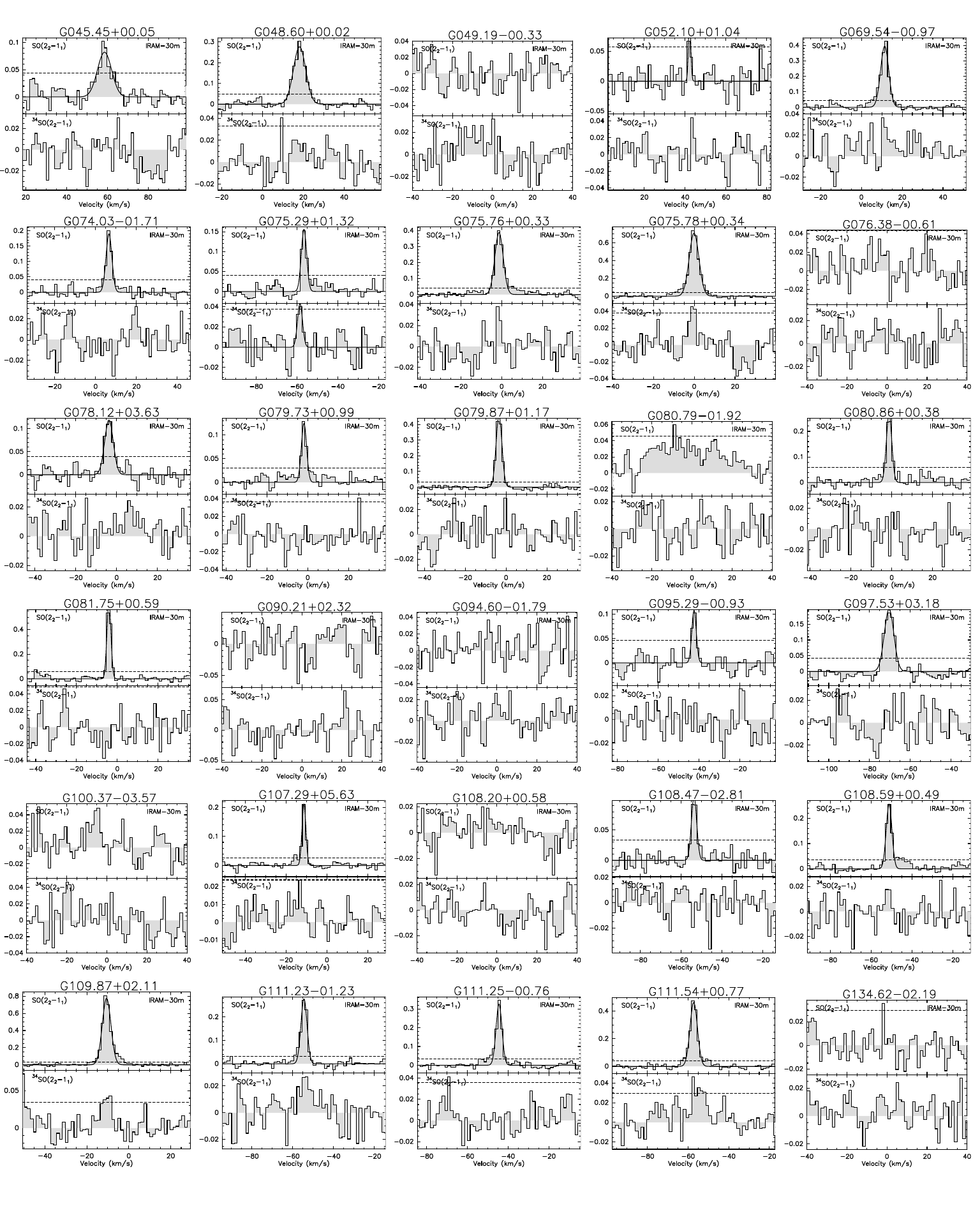}
\caption{continued.}
\end{figure*}

\addtocounter{figure}{-1}
\begin{figure*}[ht]
\centering
\includegraphics[width=0.85\textwidth, page=1]{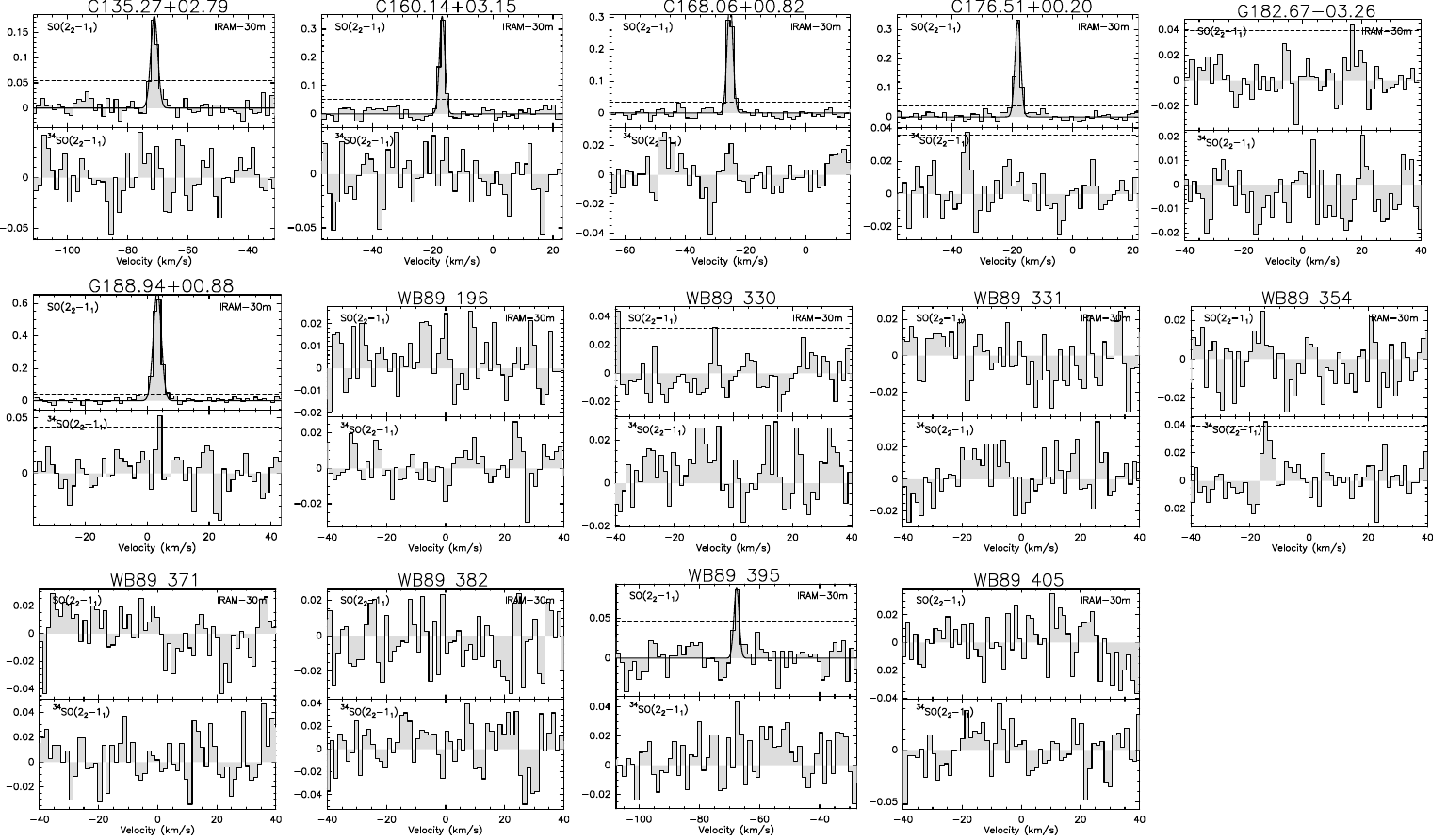}
\caption{continued.}
\end{figure*}

\begin{figure*}[ht]
\centering
\includegraphics[width=0.85\textwidth, page=1]{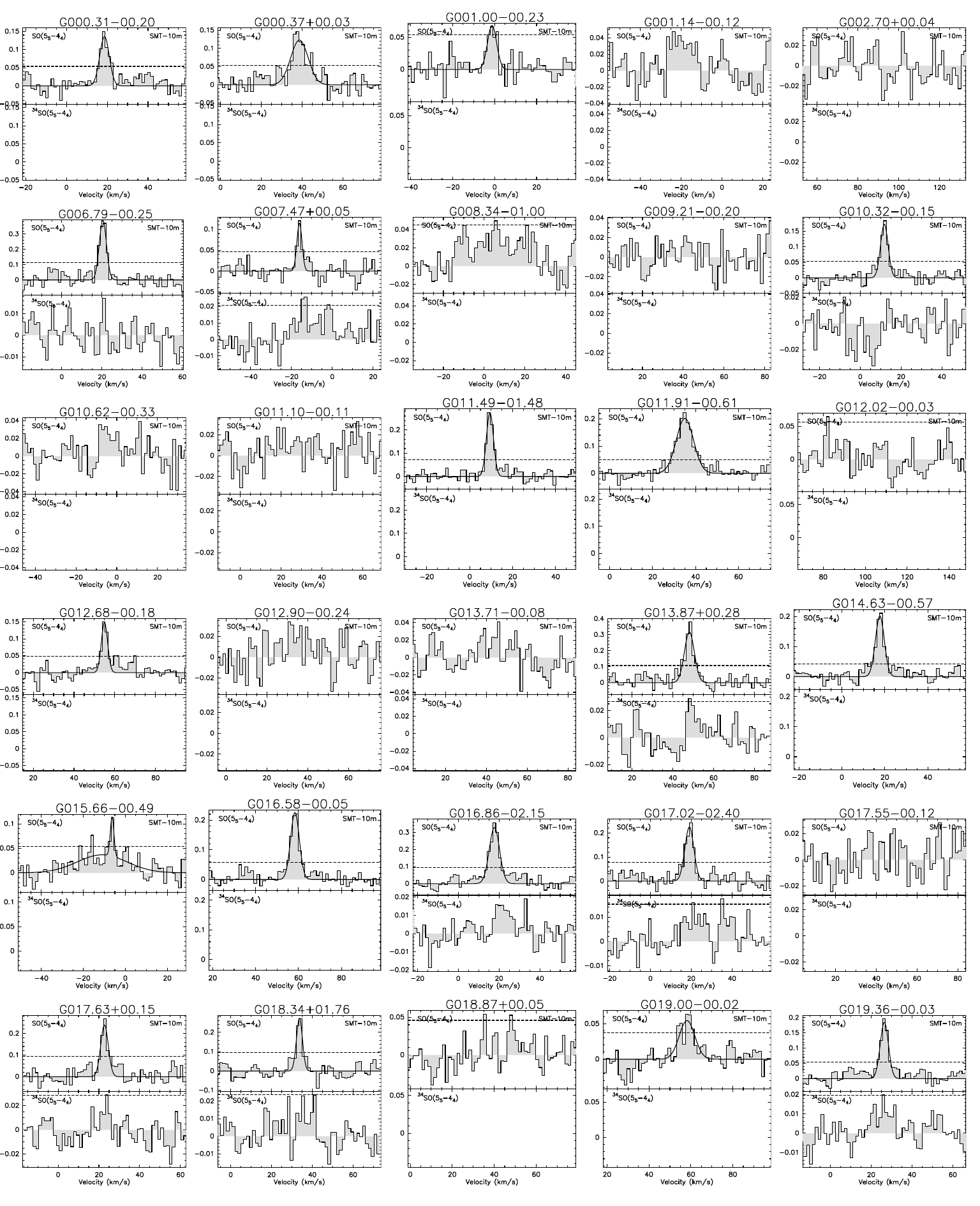}
\caption{The SMT 10 m spectra of the sources without effective measurements of SO/$^{34}$SO ($5_5-4_4$ lines).}
\label{fig:otherSMT}
\end{figure*}
\foreach \t in {2,...,4}{
\addtocounter{figure}{-1}
\begin{figure*}[ht]
\centering
\includegraphics[width=0.85\textwidth, page=\t]{other5-4}
\caption{continued.}
\end{figure*}
}
\addtocounter{figure}{-1}
\begin{figure*}[ht]
\centering
\includegraphics[width=0.85\textwidth, page=1]{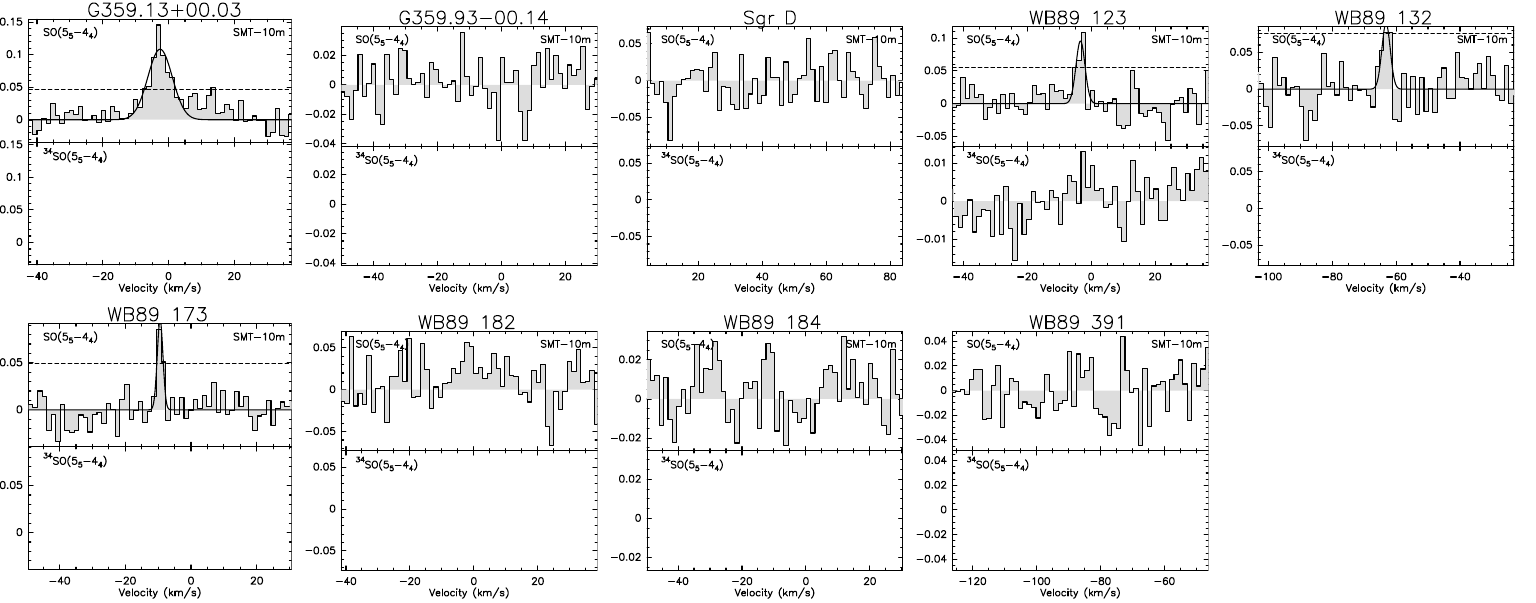}
\caption{continued.}
\end{figure*}

\clearpage
\section{Tables}\label{sec:appendixT}
\setcounter{table}{0}  
\renewcommand{\thetable}{A.\arabic{table}} 
\renewcommand*{\theHtable}{\thetable}
\setlength{\tabcolsep}{1.5mm}
\startlongtable
\begin{deluxetable}{ccccc}
\tabletypesize{\footnotesize}
\tablecaption{\label{tab:dparameter}Derived Parameters of SO.}
\tablehead{
\colhead{Source} & \colhead{$T_{rot}$} & \colhead{$N_{rot}$} & \colhead{$T_{\rm LTE}$} & \colhead{$N_{\rm LTE}$}\\
\colhead{} & \colhead{(K)} & \colhead{($\times$10$^{14}$ cm$^{-2}$)} & \colhead{(K)} & \colhead{($\times$10$^{14}$ cm$^{-2}$)}\\
\colhead{(1)} & \colhead{(2)} & \colhead{(3)} & \colhead{(4)} & \colhead{(5)}
}
\startdata
G000.37+00.03 & 13 $\pm$ 3     & 0.62 $\pm$ 0.12 & $12^{+4}_{-2}$   & $0.74^{+0.20}_{-0.20}$ \\
G002.70+00.04 & 56 $\pm$ 64    & 0.15 $\pm$ 0.13 &                  &                        \\
G007.47+00.05 & 13 $\pm$ 9     & 0.20 $\pm$ 0.04 & $11^{+3}_{-2}$   & $0.24^{+0.06}_{-0.07}$ \\
G010.62-00.33 & 20 $\pm$ 8     & 0.19 $\pm$ 0.04 &                  &                        \\
G011.49-01.48 & 17 $\pm$ 11    & 0.39 $\pm$ 0.10 & $16^{+6}_{-4}$   & $0.44^{+0.12}_{-0.10}$ \\
G012.80-00.20 & 47 $\pm$ 40    & 4 $\pm$ 3       &                  &                        \\
G012.90-00.24 & 15 $\pm$ 4     & 0.26 $\pm$ 0.05 &                  &                        \\
G013.87+00.28 & 22.8 $\pm$ 1.8 & 0.46 $\pm$ 0.09 & $22^{+11}_{-6}$  & $0.51^{+0.08}_{-0.08}$ \\
G014.63-00.57 & 14 $\pm$ 2     & 0.45 $\pm$ 0.09 & $13^{+3}_{-2}$   & $0.53^{+0.10}_{-0.10}$ \\
G016.58-00.05 & 13 $\pm$ 8     & 0.69 $\pm$ 0.17 & $11^{+2}_{-2}$   & $0.90^{+0.19}_{-0.19}$ \\
G023.00-00.41 & 17 $\pm$ 13    & 0.7 $\pm$ 0.2   & $15^{+5}_{-3}$   & $0.80^{+0.20}_{-0.19}$ \\
G023.25-00.24 & 25 $\pm$ 13    & 0.11 $\pm$ 0.03 &                  &                        \\
G023.44-00.18 & 13 $\pm$ 3     & 0.87 $\pm$ 0.18 & $13^{+3}_{-2}$   & $1.12^{+0.23}_{-0.21}$ \\
G027.36-00.16 & 17 $\pm$ 11    & 0.7 $\pm$ 0.2   & $14^{+4}_{-2}$   & $0.95^{+0.19}_{-0.18}$ \\
G028.86+00.06 & 17 $\pm$ 16    & 0.6 $\pm$ 0.2   & $14^{+4}_{-3}$   & $0.63^{+0.16}_{-0.15}$ \\
G029.86-00.04 & 86 $\pm$ 51    & 0.35 $\pm$ 0.19 & $85^{+21}_{-20}$ & $0.38^{+0.13}_{-0.12}$ \\
G031.28+00.06 & 100 $\pm$ 179  & 3 $\pm$ 4       & $46^{+0}_{-20}$  & $1.67^{+0.30}_{-0.46}$ \\
G031.58+00.07 & 15 $\pm$ 12    & 0.56 $\pm$ 0.17 & $12^{+4}_{-2}$   & $0.66^{+0.17}_{-0.18}$ \\
G032.04+00.05 & 17 $\pm$ 10    & 1.0 $\pm$ 0.3   & $13^{+3}_{-2}$   & $1.20^{+0.23}_{-0.25}$ \\
G033.64-00.22 & 197 $\pm$ 702  & 1 $\pm$ 3       &                  &                        \\
G034.39+00.22 & 18 $\pm$ 6     & 0.40 $\pm$ 0.09 &                  &                        \\
G035.14-00.73 & 15 $\pm$ 4     & 0.74 $\pm$ 0.16 & $13^{+5}_{-3}$   & $0.76^{+0.18}_{-0.17}$ \\
G037.43+01.51 & 70 $\pm$ 88    & 2 $\pm$ 3       &                  &                        \\
G040.62-00.13 & 11 $\pm$ 6     & 0.73 $\pm$ 0.15 & $11^{+2}_{-2}$   & $0.89^{+0.26}_{-0.25}$ \\
G043.89-00.78 & 19 $\pm$ 7     & 0.25 $\pm$ 0.06 & $16^{+5}_{-3}$   & $0.28^{+0.05}_{-0.05}$ \\
G045.45+00.05 & 15 $\pm$ 13    & 0.40 $\pm$ 0.12 & $14^{+4}_{-3}$   & $0.46^{+0.12}_{-0.11}$ \\
G049.49-00.37 & 14 $\pm$ 3     & 6.2 $\pm$ 1.3   &                  &                        \\
G052.10+01.04 & 23 $\pm$ 14    & 0.10 $\pm$ 0.03 & $20^{+7}_{-5}$   & $0.12^{+0.02}_{-0.02}$ \\
G073.65+00.19 & 12 $\pm$ 9     & 0.28 $\pm$ 0.06 & $11^{+3}_{-2}$   & $0.33^{+0.10}_{-0.09}$ \\
G074.03-01.71 & 16 $\pm$ 9     & 0.38 $\pm$ 0.10 & $14^{+4}_{-3}$   & $0.46^{+0.10}_{-0.12}$ \\
G076.38-00.61 & 10 $\pm$ 3     & 0.4 $\pm$ 0.3   &                  &                        \\
G078.12+03.63 & 29 $\pm$ 29    & 0.5 $\pm$ 0.3   & $22^{+8}_{-6}$   & $0.46^{+0.08}_{-0.07}$ \\
G079.73+00.99 & 16 $\pm$ 12    & 0.22 $\pm$ 0.07 & $14^{+4}_{-2}$   & $0.26^{+0.06}_{-0.06}$ \\
G080.86+00.38 & 17 $\pm$ 16    & 0.51 $\pm$ 0.19 & $15^{+5}_{-3}$   & $0.58^{+0.15}_{-0.14}$ \\
G094.60-01.79 & 6.8 $\pm$ 1.4  & 0.7 $\pm$ 0.3   &                  &                        \\
G095.29-00.93 & 15 $\pm$ 12    & 0.16 $\pm$ 0.05 & $12^{+4}_{-2}$   & $0.20^{+0.05}_{-0.06}$ \\
G097.53+03.18 & 17 $\pm$ 14    & 0.7 $\pm$ 0.2   & $15^{+5}_{-3}$   & $0.76^{+0.19}_{-0.18}$ \\
G107.29+05.63 & 20 $\pm$ 7     & 0.31 $\pm$ 0.08 & $19^{+10}_{-4}$  & $0.35^{+0.07}_{-0.08}$ \\
G108.18+05.51 & 24 $\pm$ 7     & 0.23 $\pm$ 0.05 &                  &                        \\
G108.47-02.81 & 12 $\pm$ 7     & 0.18 $\pm$ 0.04 & $11^{+3}_{-2}$   & $0.22^{+0.05}_{-0.06}$ \\
G111.25-00.76 & 17 $\pm$ 12    & 0.7 $\pm$ 0.2   & $16^{+5}_{-3}$   & $0.78^{+0.20}_{-0.18}$ \\
G135.27+02.79 & 16 $\pm$ 10    & 0.29 $\pm$ 0.08 & $13^{+3}_{-2}$   & $0.37^{+0.07}_{-0.07}$ \\
G160.14+03.15 & 78 $\pm$ 60    & 1.2 $\pm$ 0.8   & $82^{+19}_{-22}$ & $1.31^{+0.50}_{-0.44}$ \\
G168.06+00.82 & 52 $\pm$ 11    & 0.9 $\pm$ 0.2   & $53^{+18}_{-23}$ & $0.99^{+0.42}_{-0.35}$ \\
G176.51+00.20 & 18 $\pm$ 6     & 0.46 $\pm$ 0.11 &                  &                        \\
G188.94+00.88 & 34 $\pm$ 21    & 1.8 $\pm$ 0.8   &                  &                        \\
G196.45-01.67 & 17 $\pm$ 6     & 0.18 $\pm$ 0.04 &                  &                        \\
Sgr D         & 11 $\pm$ 3     & 0.14 $\pm$ 0.03 &                  &                        \\
W33           & 17 $\pm$ 4     & 2.5 $\pm$ 0.5   &                  &                        \\
WB89 123      & 10.2 $\pm$ 1.3 & 0.24 $\pm$ 0.05 &                  &                        \\
WB89 132      & 10.4 $\pm$ 1.3 & 0.20 $\pm$ 0.04 &                  &                        \\
WB89 171      & 13 $\pm$ 2     & 1.5 $\pm$ 0.3   &                  &                        \\
WB89 173      & 11.1 $\pm$ 1.5 & 0.13 $\pm$ 0.03 &                  &                        \\
WB89 395      & 11 $\pm$ 2     & 0.10 $\pm$ 0.02 &                  &                        \\
\enddata
\tablecomments{Column (1): source name, columns (2) and (3): rotational temperature and column density obtain from rotation diagram method, column (4) and (5): rotational temperature and column density obtain from ``best-fit" model (see details in Section \ref{sec:tau}).}
\end{deluxetable}

\end{appendix}

\end{CJK*}
\end{document}